\documentclass[journal]{IEEEtran}

\usepackage{amsmath,amssymb,graphicx,color,textcomp,bm,cite,stfloats}
\usepackage{algorithm,algorithmic,array,tabularx,booktabs,enumitem,tcolorbox}
\usepackage[hidelinks]{hyperref}
\newcolumntype{Y}{>{\centering\arraybackslash}X}
\tcbuselibrary{breakable}  

\providecommand{\Description}[1]{}
\makeatletter
\newtheorem{remark}{\textbf{Remark}}
\allowdisplaybreaks[1]

\begin{document}
	
\title{CL-SEC: Cross-Layer Semantic Error Correction Empowered by Language Models}

\author{Yirun Wang, Yuyang Du, Soung Chang Liew, \textit{Life Fellow, IEEE}, Yuchen Pan, Feifan Zhang, and Lihao Zhang%
\thanks{The authors are with the Department of Information Engineering, The Chinese University of Hong Kong, Hong Kong SAR, China (e-mail:\{yrwang, yuydu, soung, py025, zf024\}@ie.cuhk.edu.hk; lihaocuhk@link.cuhk.edu.hk).}
\vspace{-2em}
\thanks{Corresponding author: Soung Chang Liew.}
}

\maketitle

\begin{abstract}
    Achieving reliable communication has long been a fundamental challenge in networked systems. Semantic Error Correction (SEC) leverages the semantic understanding capabilities of language models (LMs) to perform error correction, complementing channel decoding. While promising, existing standalone SEC approaches rely solely on context captured by LMs at the application layer, ignoring the rich information available at the physical layer. To address this limitation, this paper introduces Cross-Layer SEC (CL-SEC), an LM-empowered error correction framework that integrates cross-layer information from both the physical and application layers for word correction in text communication. CL-SEC supports a broad range of channel codes equipped with soft-output decoders. Its probabilistic computations are grounded in belief propagation, with source-bit interleaving mitigating the effects of short loops in the underlying factor graph. This principled cross-layer design enables CL-SEC to substantially outperform methods that process information independently at each layer. Experiments using convolutional and low-density parity-check (LDPC) codes over different channel models demonstrate consistent improvements in bit error rate, word error rate, and semantic fidelity. 
    Importantly, unlike most semantic communication systems that focus solely on recovering the semantic meaning of transmitted messages, CL-SEC aims to reconstruct the original transmitted message verbatim, leveraging the semantic understanding capabilities of LMs for precise reconstruction.
\end{abstract}

\begin{IEEEkeywords}
Error correction, language models, semantic communication, semantic error correction, cross-layer error correction.  
\end{IEEEkeywords}

\section{Introduction}\label{sec:intro}
Error correction has been an enduring research topic in communication systems. While demanding high reliability, communication systems remain inherently susceptible to transmission errors. Classical channel coding approaches introduce controlled redundancy into transmitted data to enable forward error correction (FEC) at the receiver \cite{roth2006introduction, Ryan2009Channel}. These techniques allow the receiver to detect and correct transmission errors at the physical layer by exploiting the imposed constraints within the encoded message. While these physical-layer methods have been proven highly effective for error correction and serve as the cornerstone of modern communication systems, they may still encounter unrecoverable errors under highly adverse channel conditions. Consequently, the quest for more robust error correction methods has remained an active and ongoing focus of the research community.

The rapid development of language models (LMs) ushers in a new approach to tackling the error-correction problem. Leveraging semantic understanding acquired from massive training datasets, LMs can identify corrupted text data by recognizing linguistic patterns and revise it based on contextual coherence. Unlike conventional physical-layer methods, which rely on signal-level information, LMs exploit semantic and linguistic patterns to correct errors in a way that reflects the logic and structure of human communication. This provides a complementary approach to traditional FEC by addressing errors at the application layer through a deeper understanding of language.

\begin{figure*}[htbp]
       \centering
       \includegraphics[width=0.815\linewidth]{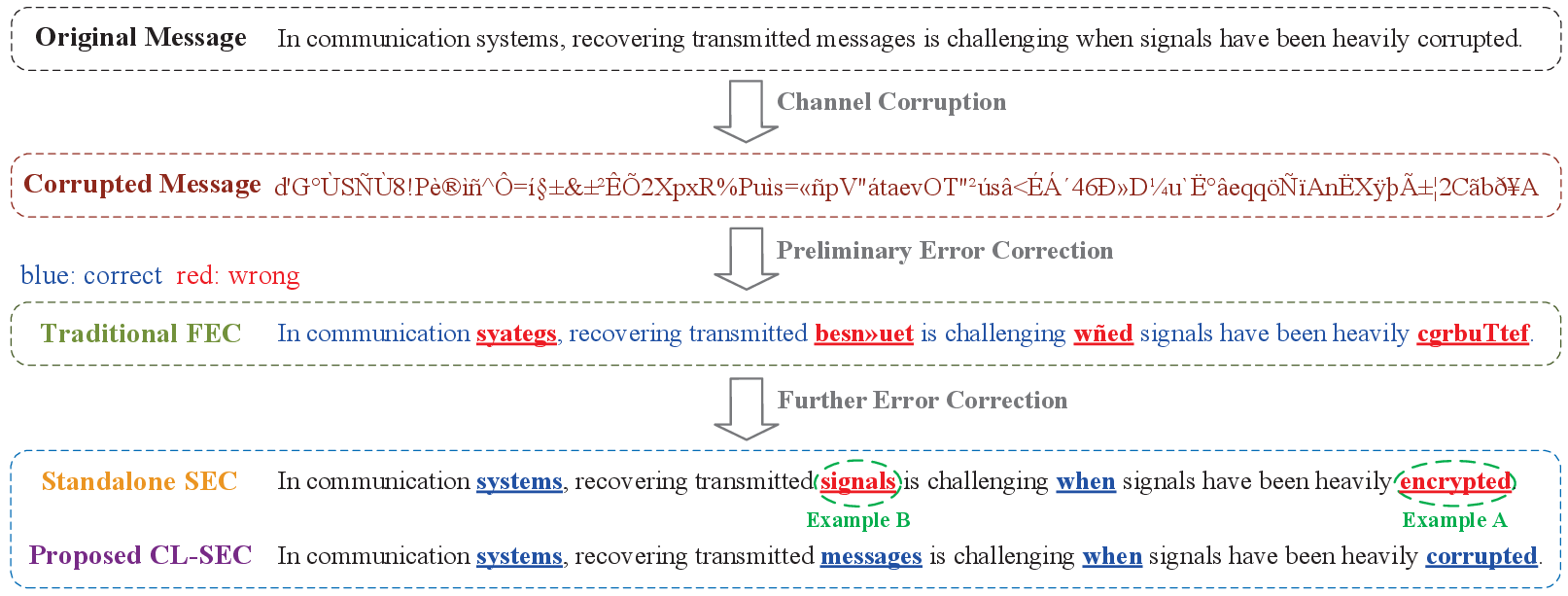} 
       \caption{Examples of error correction via different approaches.}
       \Description{A diagram showing error correction through different approaches}
       \label{fig:example}
\end{figure*}

To illustrate, consider Fig.~\ref{fig:example}, where a corrupted message is received (see the second box). After FEC at the physical layer, some errors may persist if the message has been severely corrupted (see the third box). Semantic Error Correction (SEC) leverages LMs to address these residual errors. For example, an LM can inspect the message, identify a corrupted word such as ``{\it syategs}'', and revise it to the correct word ``{\it systems}'' (see ``Standalone SEC'' in the fourth box). This standalone SEC approach---utilizing LMs for error correction---is a basic SEC method performed after FEC, and it provides an additional layer of error correction to handle remaining errors.

To motivate our approach, let us analyze the limitations of standalone SEC following FEC. Operating exclusively at the application layer, standalone SEC may still encounter problems in recovering the transmitted message under severely corrupted receptions. In particular, standalone SEC may suffer from two key shortcomings: \textbf{\textit{semantic inaccuracy}} and \textbf{\textit{lexical imprecision}}.
\begin{itemize}[leftmargin=*, label=$\bullet$]
    \item \textbf{Semantic Inaccuracy:} Standalone SEC may produce outputs that are semantically plausible but factually incorrect. For instance, in Example A (see ``Standalone SEC'' in the last box in Fig.~\ref{fig:example}), the recovered word ``{\it encrypted}'' is contextually coherent with the corrupted sentence in the third box but fails to capture the intended meaning of the original word ``{\it corrupted}''.
    \item \textbf{Lexical Imprecision:} Standalone SEC can preserve semantic meaning but fails to maintain lexical accuracy. For example, in Example B, the recovered phrase ``{\it transmitted signals}'' aligns semantically with the original message but lexically deviates from the original term ``{\it messages}''. Such deviations can lead to critical misunderstandings in contexts requiring verbatim accuracy, such as protocol specifications or technical documentation.
\end{itemize}

To address these limitations and unlock the full potential of LMs for error correction, we propose Cross-Layer Semantic Error Correction (CL-SEC). It operates after FEC to refine error correction. Unlike standalone SEC, which relies solely on the post-FEC decoded text and aims to recover the semantic meaning, CL-SEC achieves {\bf verbatim text recovery} even under severe corruption by integrating physical-layer FEC probabilities with application-layer semantic probabilities from LMs. The main contributions of this work are summarized as follows:
\begin{itemize}[leftmargin=*, label=$\bullet$]
    \item \textbf{CL-SEC Framework:} We develop CL-SEC with words as the basic correction units. The framework further incorporates context-based punctuation restoration to reconstruct complete messages.
    The framework is compatible with various soft-output channel decoders and language models, allowing flexible choices of channel codes and models to accommodate diverse communication conditions. 
    \item \textbf{Reliability-Aware Cross-Layer Word Correction:} 
    We introduce a reliability-aware correction mechanism that preserves reliable word estimates while refining unreliable ones. By combining the decoder-derived physical-layer probabilities with the LM-derived application-layer probabilities, the mechanism enables principled cross-layer word correction.
    \item \textbf{BP-Based Foundation and Interleaving Design:} 
    We formulate a factor graph for CL-SEC and derive its product-form probability computations based on belief propagation (BP). Although these computations are exact on loop-free graphs under the ideal factorized model, the CL-SEC factor graph generally contains loops. We therefore introduce source-bit interleaving to mitigate short-loop effects by increasing loop lengths or separating the words involved, thereby weakening dependencies and improving the accuracy of the product-form approximation.
    \item \textbf{Comprehensive Evaluation and Benchmarking:} We evaluate CL-SEC with convolutional and LDPC codes over AWGN and Rayleigh fading channels using bit error rate, word error rate, character error rate, and message fidelity metrics. The results demonstrate that the principled integration of semantic and physical-layer information substantially improves conventional channel decoding and methods that process information individually at each layer, enabling more accurate verbatim message recovery.
\end{itemize}


\section{System Architecture}\label{sec:architecture}

\begin{figure*}[htbp]
    \centering
    \includegraphics[width=1\linewidth]{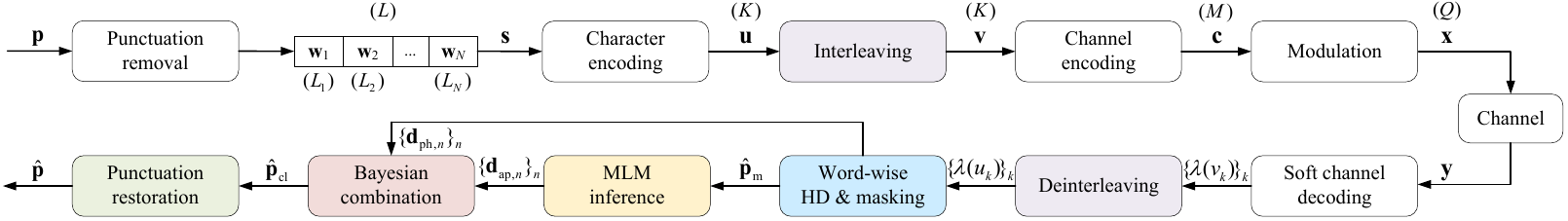} 
    \caption{The communication system with the CL-SEC framework operating at the receiver. The vector dimension is indicated in parentheses.}
    \Description{Block diagram of the CL-SEC communication system}
    \label{fig:system}
\end{figure*}

We consider the transmission of a frame containing a text message. Figure~\ref{fig:system} shows the block diagram of the communication system with the CL-SEC framework, using words as the basic correction units. 

At the transmitter, we first remove all punctuation and spaces from the original message $\mathbf{p}$. There are $N$ words left, denoted by $\mathbf{w}_n,\,n\in\mathcal{N}\triangleq\{1,\ldots,N\}$. Let $L_n$ be the number of letters in $\mathbf{w}_n$ (i.e., the word length). Concatenating all words yields a character sequence $\mathbf{s}=(\mathbf{w}_1,\ldots,\mathbf{w}_N)=(s_1,\ldots,s_L)$ of length $L = \sum_{n=1}^{N} L_n$. Information on the word lengths $\{L_n\}_{n\in\mathcal{N}}$ is included in the frame header so that the receiver knows the demarcations of words.\footnote{We employ canonical Huffman coding to compress the word-length metadata. Our evaluations in Appendix~\ref{sec:header} indicate that the header overhead is less than 10\% for 98.77\% of the passages in the corpus. Note that removing spaces and punctuation from the original text saves more bits than are required to transmit the compressed word-length metadata, resulting in a net reduction in the number of transmitted bits. The net savings can be used to introduce additional FEC redundancy, thereby providing stronger protection for the metadata.} 

For transmission, each character $s_\ell,\,\ell \in \{1, \ldots, L\}$, is converted to its 8-bit ASCII representation. Accordingly, each word $\mathbf{w}_n$ is character-encoded into $\mathbf{u}_n \in \mathbb{F}_2^{K_n}$, where $K_{n} = 8L_n$ is the number of bits within word $\mathbf{w}_n$ and $\mathbb{F}_2 \triangleq \{0,1\}$. Concatenation yields bit sequence $\mathbf{u} = (\mathbf{u}_1, \dots, \mathbf{u}_N) = (u_1, \dots u_K) \in \mathbb{F}_2^K$ with $K=8L$. 
Before channel coding, a source-bit interleaver is applied to disperse the source bits, producing the interleaved bit-stream: 
\begin{equation}
    \mathbf{v}\triangleq (v_1,\ldots,v_K) = \pi (\mathbf{u}).
\end{equation}
The readers are referred to Section~\ref{sec:justification} and Appendix~\ref{sec:supp} for details on the interleaver and the associated ablation study.

The interleaved sequence $\mathbf{v}$ is channel-encoded at code rate $R$ to produce the coded sequence $\mathbf{c}\in\mathbb{F}_2^M$. The CL-SEC framework is compatible with any channel code equipped with a soft-output decoder. In this paper, we instantiate CL-SEC using a convolutional code decoded by the BCJR algorithm \cite{Bahl1974Optimala} and a low-density parity-check (LDPC) code decoded by the belief propagation (BP) algorithm \cite{Kschischang2001Factor}. For a rate-$R$ convolutional code, each information bit is encoded into $1/R$ coded bits. For an $(N_{\mathrm{c}},K_{\mathrm{c}})$ block code, such as an LDPC code, the payload is partitioned into source blocks of $K_{\mathrm{c}}$ bits, with the final block zero-padded if necessary. The blocks are then encoded independently, and the resulting codewords are concatenated and rate-matched to achieve an effective code rate of $R$. Finally, the coded bits are  mapped to the modulated symbol sequence $\mathbf{x}=(x_1,\ldots,x_Q) \in\mathbb{C}^{Q} $. 

After transmission through the channel, the received symbol sequence is given by
\begin{equation}
    \label{eq:y}
    \mathbf{y}= h \mathbf{x}+\mathbf{n},
\end{equation}
where $h$ is the channel coefficient and $\mathbf{n}\sim\mathcal{CN}(\mathbf{0}_Q,\sigma^2\mathbf{I}_Q)$ is the complex Gaussian noise. For additive white Gaussian noise (AWGN) channels, we have $ h=1 $. For Rayleigh fading channels, $h\sim\mathcal{CN}(0,1)$, corresponding to the unit average channel power. This paper evaluates CL-SEC over both AWGN and Rayleigh fading channels. 

At the receiver, soft decoding of the received symbol sequence $\mathbf{y}$ produces bit-level log-likelihood ratios (LLRs) $\{\lambda(v_k)\}_{k\in\mathcal{K}}$ in the interleaved order, where $ \mathcal{K}\triangleq\{1,\ldots,K\} $. Soft decoding is performed using the BCJR algorithm for convolutional codes and the BP algorithm for LDPC codes. Deinterleaving restores these LLRs to the original source-bit order, yielding $\{\lambda(u_k)\}_{k\in\mathcal{K}}$. In conventional FEC decoding, bit-wise hard decision (HD) is applied directly to these LLRs to obtain the decoded bit sequence. Under severe channel conditions, however, the resulting sequence may still contain residual errors. As shown in Fig.~\ref{fig:system}, CL-SEC improves decoding reliability by combining the decoder's soft information with contextual likelihoods provided by the LM to further refine the channel decoding output. 


Specifically, CL-SEC first performs word-wise HD (WW-HD) on the deinterleaved LLRs to obtain a word estimate and an associated confidence score at each word position. Words with unreliable estimates are then masked, whereas those with reliable estimates are retained as context, forming the partially masked sequence $\hat{\mathbf{p}}_{\mathrm{m}}$ for input to the masked language model (MLM). At each masked-word position $ n $, Bayesian fusion combines the physical-layer candidate distribution $ \mathbf{d}_{\mathrm{ph},n} $ derived from the LLRs with the application-layer candidate distribution $ \mathbf{d}_{\mathrm{ap},n} $ generated by the MLM. The resulting cross-layer distribution determines the corrected word. Replacing each mask in $\hat{\mathbf{p}}_{\mathrm{m}}$ with the corresponding corrected word yields the cross-layer corrected sequence $\hat{\mathbf{p}}_{\mathrm{cl}}$. Finally, context-based punctuation restoration is applied to $\hat{\mathbf{p}}_{\mathrm{cl}}$ to produce the final recovered message $\hat{\mathbf{p}}$. Section~\ref{sec:cl-sec} describes these CL-SEC operations in detail.

\section{CL-SEC Framework}\label{sec:cl-sec} 

\subsection{Overview}\label{sec:overview}

CL-SEC uses words as its correction units. As illustrated in  Fig.~\ref{fig:system}, it performs the following operations at the receiver:

\textbf{1) Soft Channel Decoding} (Section~\ref{sec:hd}): Given the received symbols $\mathbf{y}$, the soft channel decoder produces the bit LLRs $\{\lambda(v_k)\}_{k\in\mathcal{K}}$ in the interleaved order. Deinterleaving restores them to the original source-bit order, yielding $\{\lambda(u_k)\}_{k\in\mathcal{K}}$.

\textbf{2) Word-Wise HD and Masking} (Section~\ref{sec:detect}): For each word position $n\in\mathcal{N}$, CL-SEC constructs a candidate set $\mathcal{W}_n$ consisting of words whose lengths match the corresponding word-length metadata. Using the relevant bit LLRs, word-wise HD constructs a physical-layer probability distribution $\mathbf{d}_{\mathrm{ph},n}$ over these candidates and selects the most probable candidate as the initial word estimate. Word positions with confidence scores below a prescribed threshold are masked for further correction, whereas the remaining estimates are confirmed and retained as context.

\textbf{3) MLM Inference} (Section~\ref{sec:app}): Spaces are inserted between adjacent word positions to form a partially masked sequence $\hat{\mathbf{p}}_{\mathrm{m}}$. The MLM takes $\hat{\mathbf{p}}_{\mathrm{m}}$ as input and uses the confirmed words as context to infer each masked word, producing the application-layer candidate distribution $\mathbf{d}_{\mathrm{ap},n}$ at each masked position. 

\textbf{4) Bayesian Combination} (Section~\ref{sec:cl}): For each masked-word position $n$, CL-SEC Bayesian-combines the complementary physical-layer and application-layer distributions to obtain the cross-layer posterior distribution $\mathbf{d}_{\mathrm{cl},n}$. The candidate with the highest posterior probability is selected as the corrected word $\hat{\mathbf{w}}_{\mathrm{cl},n}$. Replacing each mask in $\hat{\mathbf{p}}_{\mathrm{m}}$ with its corresponding corrected word yields the cross-layer corrected sequence $\hat{\mathbf{p}}_{\mathrm{cl}}$. 

\textbf{5) Punctuation Restoration} (Section~\ref{sec:pr}): Finally, CL-SEC restores punctuation in the cross-layer corrected sequence $\hat{\mathbf{p}}_{\mathrm{cl}}$ based on its context, producing the final recovered message $\hat{\mathbf{p}}$.

After presenting the individual operations in Section~\ref{sec:hd} to Section~\ref{sec:pr}, we summarize the overall CL-SEC operations in Algorithm~\ref{alg:clsec}. Section~\ref{sec:justification} then provides a factor-graph interpretation and BP-based justification for the probability computations used in CL-SEC.

\subsection{Soft Channel Decoding} \label{sec:hd}

Before delving into CL-SEC error correction, let us first review the soft channel decoding, which provides preliminary decoding and generates the bit-level LLRs required by CL-SEC (Fig.~\ref{fig:system}). 
Such LLRs are readily available from channel codes equipped with soft-output channel decoders. This paper employs the convolutional code with a BCJR decoder \cite{Bahl1974Optimala} and the LDPC code with a BP decoder \cite{Kschischang2001Factor}. However, CL-SEC is compatible with other channel codes with a soft-output decoder. The BCJR algorithm performs exact sum-product BP on the loop-free trellis factor graph of a convolutional code, whereas standard LDPC decoding performs iterative sum-product BP on a Tanner graph that generally contains loops.

Given $ \mathbf{y} $, the soft channel decoder outputs bit-wise posterior probabilities $P(v_k|\mathbf{y})$ and the soft LLR $\lambda (v_k)$ for each source bit $v_k$, $k\in\mathcal{K}$. Deinterleaving yields $P(u_k|\mathbf{y})$ and $ \lambda (u_k) $:
\begin{equation}\label{eq:llr}
    \lambda (u_k) \triangleq \log\left(\frac{ P (\mathbf{y}|u_k=0) }
     {P (\mathbf{y}|u_k=1 ) }\right) 
    \stackrel{(a)}{=} 
    \log\left(\frac{ P (u_k=0|\mathbf{y} ) }
    { P (u_k=1|\mathbf{y} ) }\right), 
\end{equation} 
where $(a)$ follows from the uniform bit-prior assumption adopted by the channel decoder, i.e., $P(u_k = 0) = P(u_k = 1) = 1/2$. 

For the HD stage of CL-SEC, decisions are made at the word level. The conventional bit-wise HD scheme below serves solely as a benchmark for comparison and is not used within the CL-SEC framework.

\textit{Benchmarking Bit-Wise HD (BW-HD):}
As a physical-layer benchmark, BW-HD makes an independent hard decision on each post-FEC bit:
\begin{equation}\label{eq:u_hd}
    \hat{u}_{\mathrm{bit},k} = \underset{u \in \mathbb{F}_2}{\arg\max} \, P \big(u_k=u \,|\, \mathbf{y}\big), \; k \in \mathcal{K}.
\end{equation}
Mapping the bit sequence $ (\hat{u}_{\mathrm{bit},1}, \dots, \hat{u}_{\mathrm{bit},K} )$ to characters and segmenting the resulting character sequence into words yield $(\hat{\mathbf{w}}_{\mathrm{bit},1}, \dots, \hat{\mathbf{w}}_{\mathrm{bit},N} )$, where 
$\{\hat{\mathbf{w}} _{\mathrm{bit},n}\}_{n\in\mathcal{N}}$ are the BW-HD word estimates. 

As BW-HD makes independent decisions on individual bits without enforcing word-level constraints, a resulting BW-HD word estimate may not correspond to a valid word. In contrast, the word-wise HD (WW-HD) scheme used in CL-SEC incorporates word-level constraints, as described next. 

\subsection{Word-Wise HD and Masking}\label{sec:detect}

Let $\mathcal{W}$ be the complete vocabulary. Using the word-length metadata, we construct, for each position $n \in \mathcal{N}$, a candidate set $ \mathcal{W}_n = \{ \mathbf{w}\in\mathcal{W}: |\mathbf{w}| = L_n \} $ containing all vocabulary words of length $L_n$. Let $S_n = |\mathcal{W}_n|$ denote the number of candidates and define $ \mathcal{S}_n = \{1, \dots, S_n\} $. Each candidate $\tilde{\mathbf{w}}_{n,s} \in \mathcal{W}_n,\,s \in \mathcal{S}_n $, is converted into its ASCII representation $\tilde{\mathbf{u}}_{n,s} = (\tilde{u}_{n,s} (1), \dots, \tilde{u}_{n,s}(K_n)) \in \mathbb{F}_2^{K_n}$, where $K_n$ is the number of source bits corresponding to the word at position $n$. We further define $\tilde{K}_n = \sum_{j=1}^{n-1} K_j$ as the total number of source bits preceding the word at position $n$. 

For each candidate $\tilde{\mathbf{w}}_{n,s}\in\mathcal{W}_n$ at word position $n$, we compute a score from the corresponding bit-wise posterior probabilities as
\begin{equation}\label{eq:wwscore}
\phi_n( \tilde{\mathbf{w}}_{n,s} )=\prod_{k=1}^{K_n} P\big( u_{\tilde{K}_n+k} = \tilde{u}_{n,s} (k) \,\big|\,\mathbf{y} \big), \;n\in\mathcal{N}, s \in \mathcal{S}_n.
\end{equation}
Define the physical-layer distribution over the candidate set $\mathcal{W}_n$ as 
$\mathbf{d}_{\mathrm{ph},n} = (d_{\mathrm{ph},n}(\tilde{\mathbf{w}}_{n,1}) , \ldots, d_{\mathrm{ph},n}(\tilde{\mathbf{w}}_{n,S_n}) )$. Since not every length-compatible bit sequence corresponds to a valid word, $ \sum_{\tilde{\mathbf{w}}_{n,s}\in\mathcal{W}_n}\phi_n( \tilde{\mathbf{w}}_{n,s} ) \neq 1$. The $s$th element of $ \mathbf{d}_{\mathrm{ph},n} $ is thus found from normalization:
\begin{equation}\label{eq:ph_distribution}
    d_{\mathrm{ph},n} (\tilde{\mathbf{w}}_{n,s}) = 
    \frac{\phi_n( \tilde{\mathbf{w}}_{n,s} )}
    {\sum_{j=1}^{S_n} \phi_n( \tilde{\mathbf{w}}_{n,j} )},
    \; n\in\mathcal{N}, s \in \mathcal{S}_n.
\end{equation}
At each word position $n$, WW-HD selects the candidate with the highest physical-layer probability and takes the corresponding probability as its confidence:
\begin{align}
    \hat{\mathbf{w}}_{\mathrm{ph},n}&=
    \underset{\tilde{\mathbf{w}}_{n,s}\in\mathcal{W}_n}{\arg\max}\,
    d_{\mathrm{ph},n} ( \tilde{\mathbf{w}}_{n,s} ), 
    \label{eq:w_ph}\\
    d_{\mathrm{ph},n}^{\star} &=\max_{\tilde{\mathbf{w}}_{n,s}\in\mathcal{W}_n} d_{\mathrm{ph},n}  (\tilde{\mathbf{w}}_{n,s} ),\;
    n\in\mathcal{N}. 
    \label{eq:d_ph*}
\end{align}

Given a threshold $\tau\in[0,1]$, we define the set of masked-word positions as
\begin{equation}\label{eq:Nm}
    \mathcal{N}_{\mathrm{m}}=\big\{n\in\mathcal{N}: d_{\mathrm{ph},n}^{\star}<\tau\big\}.
\end{equation}
Each unreliable word $ \hat{\mathbf{w}}_{\mathrm{ph},n},\,n\in\mathcal{N}_{\mathrm{m}} $ is replaced by a mask symbol (e.g., \textless mask\textgreater{}) for subsequent correction. The remaining words, indexed by $ n\in\mathcal{N}\setminus\mathcal{N}_{\mathrm{m}} $, are retained and left unmasked. Further inserting spaces between adjacent word positions yields $ \hat{\mathbf{p}}_{\mathrm{m}} $, the sequence of unmasked words and masks. For example, $ \hat{\mathbf{p}}_{\mathrm{m}} = $ ``There is a beach with palm \textless mask\textgreater{} and clear blue water''. We next describe MLM inference and the construction of the application-layer candidate distributions used by CL-SEC.

\subsection{MLM Inference}\label{sec:app}
Pretrained masked language models (MLMs), such as mmBERT \cite{Marone2025mmBERT2}, enable word correction by exploiting the application-layer linguistic context.\footnote{We adopt mmBERT \cite{Marone2025mmBERT2} as the pretrained MLM in our implementation. However, CL-SEC is also compatible with other pretrained models capable of mask infilling, such as BART \cite{Lewis2020BART}. Appendix~\ref{sec:supp} evaluates mmBERT-based and BART-based CL-SEC with both convolutional and LDPC codes. Simulations show that mmBERT-based CL-SEC achieves better correction performance under the considered settings.}
For example, given the sequence ``There is a beach with palm \textless mask\textgreater{} and clear blue water'', a pretrained MLM may assign the highest probability to the word ``trees'', selecting it as the prediction at the masked position.


Given the input sequence $\hat{\mathbf{p}}_{\mathrm{m}}$, the MLM generates a probability distribution over the vocabulary at each masked-word position:\footnote{Words at this stage can still be heavily corrupted, making accurate punctuation restoration challenging. Fortunately, the MLM still makes effective inference even when punctuation is absent from the input, as demonstrated by our simulations. Punctuation will be restored based on context after accurate CL-SEC word correction (Section~\ref{sec:pr}).}
\begin{equation}\label{eq:theta}
    \theta_n(\mathbf{w}) = 
    P\big(\mathbf{w}_n=\mathbf{w} \,\big|\, \hat{\mathbf{p}}_{\mathrm{m}},\mathrm{MLM} \big),\,n\in\mathcal{N}_{\mathrm{m}},\mathbf{w}\in\mathcal{W},
\end{equation}
where ``MLM'' represents the adoption of a specific instance of MLM, and
$\sum_{\mathbf{w}\in\mathcal{W}}\theta_n(\mathbf{w})=1$.

The distribution in \eqref{eq:theta} is defined over all words in the vocabulary and is not restricted to words of length $L_n$.
We retain only the probabilities assigned to candidates in
$\mathcal{W}_n\subset\mathcal{W}$ and renormalize them to obtain the application-layer
candidate distribution  
$\mathbf{d}_{\mathrm{ap},n}=(d_{\mathrm{ap},n}(\tilde{\mathbf{w}}_{n,1}),\ldots,d_{\mathrm{ap},n}(\tilde{\mathbf{w}}_{n,S_n}))$. 
Its $s$th entry is given by
\begin{equation}\label{eq:ap_prob}
    d_{\mathrm{ap},n}(\tilde{\mathbf{w}}_{n,s})=
    \frac{\theta_n(\tilde{\mathbf{w}}_{n,s})}
    {\sum_{j=1}^{S_n}\theta_n(\tilde{\mathbf{w}}_{n,j})},
     \;n\in\mathcal{N}_{\mathrm{m}}, s\in\mathcal{S}_n.
\end{equation}

\textit{Standalone MLM-Based SEC Baseline:} Before introducing the dual-layer fusion in CL-SEC, we define standalone MLM-based SEC schemes that use only application-layer information. Such schemes can be divided into two types according to whether the received word-length information is used. A basic standalone scheme, such as that illustrated in the fourth box of Fig.~\ref{fig:example}, allows the MLM to select from the entire vocabulary. Consequently, the corrected word may not have the same length as the transmitted word, as illustrated by the prediction “signals” in Fig.~\ref{fig:example}.


For a fair comparison with CL-SEC in Section~\ref{sec:sim}, we consider a stronger standalone baseline, referred to as the MLM scheme, that incorporates word-length information by restricting the MLM decision to the candidate set $\mathcal{W}_n$. Specifically, for each $n\in\mathcal{N}_{\mathrm{m}}$, the corrected word is determined solely from the application-layer distribution $\mathbf{d}_{\mathrm{ap},n}$ as
\begin{equation}\label{eq:app_corr}
    \hat{\mathbf{w}}_{\mathrm{ap},n} = \underset{\tilde{\mathbf{w}}_{n,s} \in \mathcal{W}_n}{\arg\max} \, d_{\mathrm{ap},n}(\tilde{\mathbf{w}}_{n,s}), \; n \in \mathcal{N}_{\mathrm{m}}. 
\end{equation}
Like WW-HD, the MLM scheme restricts its decision to $\mathcal{W}_n$ and thus enforces the available word-level constraints. Unlike WW-HD, however, it ranks the candidates using the linguistic context captured by the MLM for masked-word correction.

Unlike CL-SEC, the MLM scheme uses only $\mathbf{d}_{\mathrm{ap},n}$ when correcting a masked word and does not exploit the corresponding physical-layer distribution $\mathbf{d}_{\mathrm{ph},n}$. As shown in Section~\ref{sec:sim}, discarding this physical-layer information limits its correction performance. CL-SEC instead combines $\mathbf{d}_{\mathrm{ph},n}$ and $\mathbf{d}_{\mathrm{ap},n}$ to determine the corrected word at each position $n\in\mathcal{N}_{\mathrm{m}}$, as elaborated next.


\begin{figure}[t]
    \centering
    \includegraphics[width=1\linewidth]{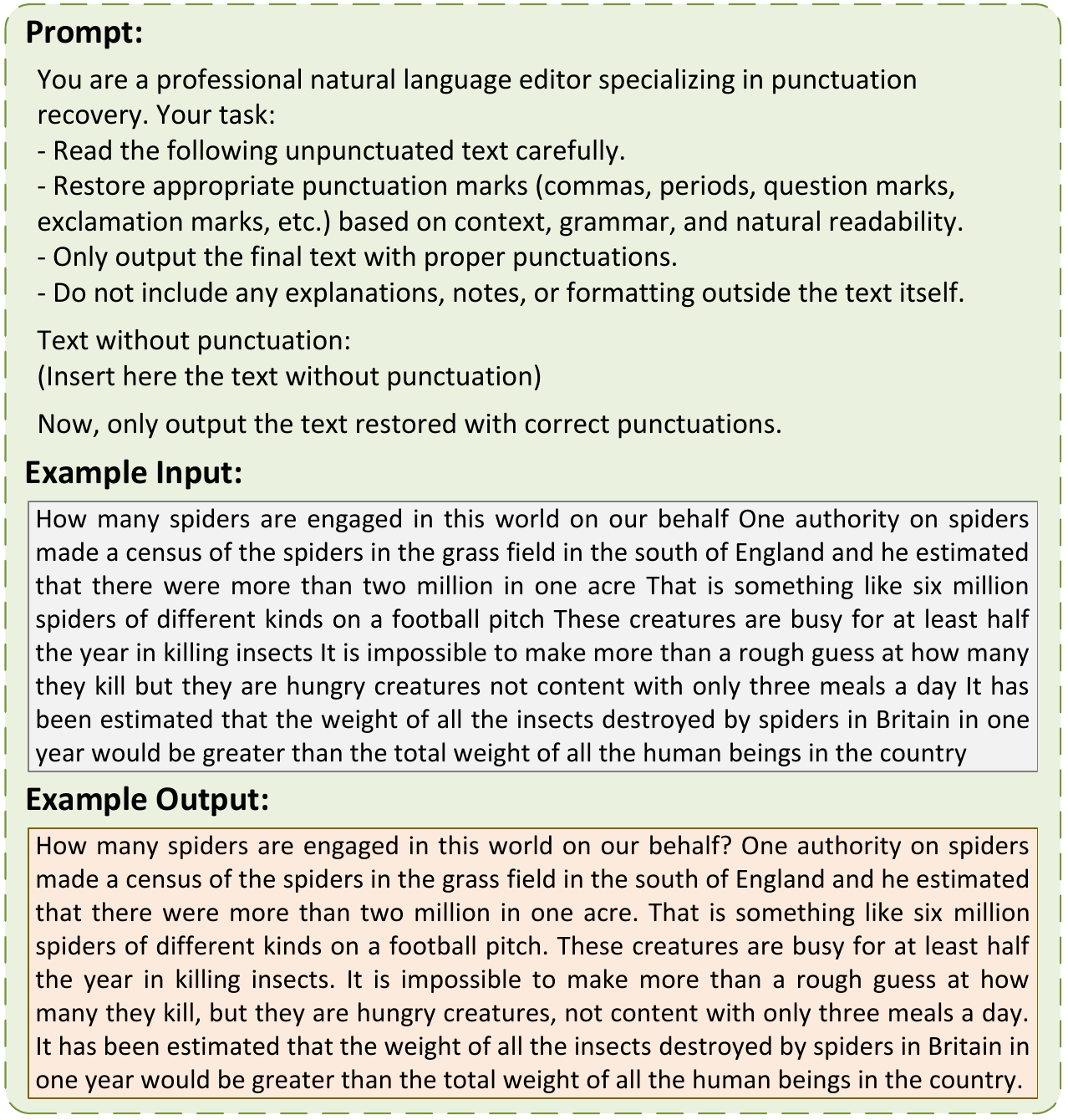}
    \caption{Prompt, example input, and example output of punctuation restoration.}
    \Description{Demonstration of punctuation restoration process}
    \label{fig:punctuation}
\end{figure}

\subsection{Bayesian Combination and Cross-Layer Word Correction}\label{sec:cl}

As described in Sections~\ref{sec:detect} and \ref{sec:app}, at each masked position $n \in \mathcal{N}_{\mathrm{m}}$, we obtain two probability distributions  over the candidate set $\mathcal{W}_n$: the physical-layer distribution $\mathbf{d}_{\mathrm{ph},n}$ derived from the decoder's LLRs, and the application-layer distribution $\mathbf{d}_{\mathrm{ap},n}$ produced by the MLM. We fuse these two distributions using the following normalized product rule:
\begin{equation}\label{eq:cross_prob}
d_{\mathrm{cl},n}(\tilde{\mathbf{w}}_{n,s}) = \frac { d_{\mathrm{ph},n}(\tilde{\mathbf{w}}_{n,s})d_{\mathrm{ap},n}(\tilde{\mathbf{w}}_{n,s}) }
{\sum_{j=1}^{S_n} d_{\mathrm{ph},n}(\tilde{\mathbf{w}}_{n,j})d_{\mathrm{ap},n}(\tilde{\mathbf{w}}_{n,j})},\; n\in\mathcal{N}_{\mathrm{m}}, s \in \mathcal{S}_n. 
\end{equation}
The resulting cross-layer probability distribution is $\mathbf{d}_{\mathrm{cl},n} = (d_{\mathrm{cl},n}(\tilde{\mathbf{w}}_{n,1}) , \ldots, d_{\mathrm{cl},n}(\tilde{\mathbf{w}}_{n,S_n}) )$. 
The candidate with the highest cross-layer probability is then selected as
\begin{equation}\label{eq:cross_corr}
\hat{\mathbf{w}}_{\mathrm{cl},n}=\underset{\tilde{\mathbf{w}}_{n,s}\in\mathcal{W}_n}{\arg\max}\ d_{\mathrm{cl},n}(\tilde{\mathbf{w}}_{n,s}),\; n\in\mathcal{N}_{\mathrm{m}}.
\end{equation}
Replacing each mask in $\hat{\mathbf{p}}_{\mathrm{m}}$ with $\{\hat{\mathbf{w}}_{\mathrm{cl},n}\}_{n \in \mathcal{N}_{\mathrm{m}}}$ accordingly, we obtain the cross-layer corrected sequence $\hat{\mathbf{p}}_{\mathrm{cl}}$. 

Like the MLM scheme, CL-SEC exploits word-level constraints and contextual linguistic information when correcting unreliable words. In addition, it incorporates physical-layer evidence from the decoder's LLRs to distinguish among linguistically plausible candidates. This evidence can also help CL-SEC better select the appropriate letter-case variant when $\mathcal{W}_n$ contains lowercase, uppercase, or initial-capitalized versions of the same word. By fusing physical-layer evidence with MLM-derived contextual probabilities, CL-SEC improves the ability to recover the transmitted message verbatim.

\begin{algorithm}[t]
\caption{The CL-SEC framework.}
\renewcommand{\algorithmicrequire}{\textbf{Input:}}
\renewcommand{\algorithmicensure}{\textbf{Output:}}
\label{alg:clsec}
\begin{algorithmic}[1]
    \REQUIRE Received symbols $\mathbf{y}$, word lengths $\{L_n\}_{n\in\mathcal{N}}$, vocabulary $\mathcal{W}$, and interleaver $\pi$. 
    \STATE Use $\mathbf{y}$ to obtain $\{\lambda(v_k)\}_{k\in\mathcal{K}}$ via soft channel decoder.
    \STATE Deinterleave to obtain $\{\lambda(u_k)\}_{k\in\mathcal{K}}$ in natural order. 
    \FOR{each $n\in\mathcal{N}$}
        \STATE Construct $\mathcal{W}_n$ from vocabulary words of length $L_n$.
        \STATE Compute distribution $\mathbf{d}_{\mathrm{ph},n}$ using \eqref{eq:wwscore}--\eqref{eq:ph_distribution}.
        \STATE Determine $\hat{\mathbf{w}}_{\mathrm{ph},n}$ and $d_{\mathrm{ph},n}^{\star}$ using \eqref{eq:w_ph}--\eqref{eq:d_ph*}.
    \ENDFOR
    \STATE Determine the masked-word positions $\mathcal{N}_{\mathrm{m}}$ using \eqref{eq:Nm}. 
    \STATE Mask $\{\hat{\mathbf{w}}_{\mathrm{ph},n}\} _{n\in\mathcal{N}_{\mathrm{m}}}$ and insert spaces to form $\hat{\mathbf{p}}_{\mathrm{m}}$.
    \STATE Feed $\hat{\mathbf{p}}_{\mathrm{m}}$ to the MLM and compute $\mathbf{d}_{\mathrm{ap},n}$ via \eqref{eq:theta}--\eqref{eq:ap_prob}. 
    \FOR{each $n\in\mathcal{N}_{\mathrm{m}}$}
        \STATE Compute the cross-layer distribution $\mathbf{d}_{\mathrm{cl},n}$ using \eqref{eq:cross_prob}.
        \STATE Select the cross-layer correction $\hat{\mathbf{w}}_{\mathrm{cl},n}$ using \eqref{eq:cross_corr}.
    \ENDFOR
    \STATE Replace masks in $\hat{\mathbf{p}}_{\mathrm{m}}$ with $\{\hat{\mathbf{w}}_{\mathrm{cl},n}\} _{n\in\mathcal{N}_{\mathrm{m}}}$ to form $\hat{\mathbf{p}}_{\mathrm{cl}}$.
    \STATE Instruct Qwen3 to restore punctuation for $\hat{\mathbf{p}}_{\mathrm{cl}}$, yielding $\hat{\mathbf{p}}$.
    \ENSURE Recovered message $\hat{\mathbf{p}}$.
\end{algorithmic}
\end{algorithm}

\subsection{Punctuation Restoration}\label{sec:pr}

The cross-layer corrected sequence $\hat{\mathbf{p}}_{\mathrm{cl}}$ contains the recovered words but lacks the punctuation present in the original message $\mathbf{p}$, which may impair readability and alter the intended interpretation. As illustrated in Fig.~\ref{fig:punctuation}, we prompt Qwen3-4B \cite{Yang2025Qwen3} to restore the missing punctuation based on the linguistic context. 
The resulting sequence constitutes the final recovered message $\hat{\mathbf{p}}$.

 \begin{figure*}[t]
       \centering
       \includegraphics[width=1\linewidth]{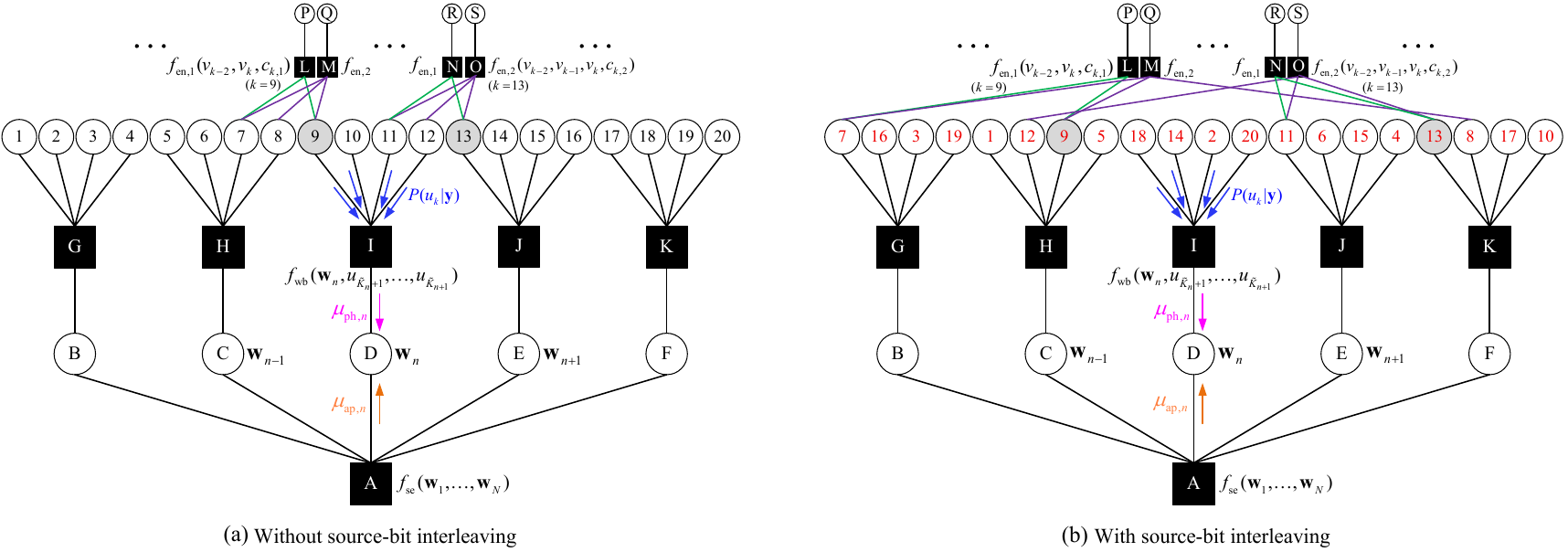} 
       \caption{A factor graph for CL-SEC assuming convolutional coding. Circles and squares represent variable nodes and factor nodes, respectively. Figures (a) and (b) show the cases without and with source-bit interleaving, respectively.}
       \Description{Factor graph showing the relationships in CL-SEC}
       \label{fig:factor_graph}
 \end{figure*}

\section{BP-Based Foundation for CL-SEC}\label{sec:justification}

\subsection{Factor Graph}

We justify the CL-SEC probability calculations from the factor-graph and BP perspectives, as illustrated in Fig.~\ref{fig:factor_graph}. To distinguish information exchanged during BP from the transmitted message in the communication system, we refer to the former as a \textbf{BP message} throughout this section.

Figures~\ref{fig:factor_graph}(a) and (b) show the CL-SEC factor graphs without and with source-bit interleaving, respectively. Circles represent variable nodes, and squares represent factor nodes. From bottom to top, the graph contains the following three types of factor nodes:

\textbf{1) Semantic Factor:} The semantic factor $f_{\mathrm{se}}(\mathbf{w}_1,\ldots,\mathbf{w}_N)$, represented by Node A, models the contextual relationships among the words in the text message. It connects the word variables, such as Nodes B to F in Fig.~\ref{fig:factor_graph}, and supplies contextual information for word inference. In CL-SEC, the corresponding factor-to-variable BP message is approximated by the masked-word distribution produced by the MLM; the high-dimensional factor $f_{\mathrm{se}}(\mathbf{w}_1,\ldots,\mathbf{w}_N)$ is therefore conceptual and is not explicitly constructed.

\textbf{2) Word-Bit Mapping Factors:}
Only certain bit combinations correspond to valid words. Nodes G to K in Fig.~\ref{fig:factor_graph} capture this relationship. Each factor $f_{\mathrm{wb}}(\mathbf{w}_n,u_{\tilde K_n+1},\ldots,u_{\tilde K_{n+1}})$ with $\tilde K_n=\sum_{j=1}^{n-1}K_j$, enforces the mapping between word $\mathbf w_n$ and its corresponding source bits:
\begin{align}
    &f_{\mathrm{wb}}\big(
        \mathbf{w}_n = \tilde{\mathbf{w}}_{n,s},u_{\tilde{K}_n+1} = b_1,\ldots,u_{\tilde{K}_{n+1}}=b_{K_n}\big) 
        \nonumber \\ 
    &\quad\;\;=\mathbb{I} \Big(
    \big(\tilde{u}_{n,s} (1), \dots, \tilde{u}_{n,s}(K_n) \big)
    = \big( b_1,\ldots,b_{K_n} \big) \Big),\label{eq:wb_mapping}
\end{align}
where $\mathbb I(\cdot)$ is the indicator function, $(\tilde u_{n,s}(1),\ldots,\tilde u_{n,s}(K_n))$ is the bit-level representation of the candidate $\tilde{\mathbf w}_{n,s}$, and $(b_1,\ldots,b_{K_n})\in\mathbb F_2^{K_n}$. For simplicity, each word is represented by only four bits in Fig.~\ref{fig:factor_graph}; words in the actual system are encoded in ASCII and have more bits.

In both Figs.~\ref{fig:factor_graph}(a) and~\ref{fig:factor_graph}(b), the source-bit variable nodes are drawn from left to right according to their positions in the original, uninterleaved source-bit sequence $\mathbf u$. Their numerical labels, however, indicate their positions in the encoder-input sequence $\mathbf v$. Without interleaving, $\mathbf v=\mathbf u$, and these two position indices coincide; with interleaving, $\mathbf v=\pi(\mathbf u)$, and the two indices are related by the interleaving permutation $\pi$.

\textbf{3) Channel-Coding Factors:} The factors $f_{\mathrm{en}}$, such as Nodes L to O, capture the channel-coding constraints between the source bits and the channel-coded bits (such as Nodes P to S in Fig.~\ref{fig:factor_graph}). Figure~\ref{fig:factor_graph} assumes a convolutional code; a similar justification applies to block codes such as LDPC codes, as discussed shortly.

At encoding step $k$, the convolutional encoder with memory length $\nu$ generates $1/R$ coded bits, denoted by $c_{k,1},\ldots,c_{k,1/R}$, each as a binary linear combination of a subset of $\{v_{k-\nu},\ldots,v_k\}$ specified by its generator polynomial. For each $j=1,\ldots,1/R$, the corresponding coding constraint is represented by the indicator factor $f_{\mathrm{en},j}(v_{k-\nu},\ldots,v_k,c_{k,j})$, which equals one when the input and coded bits satisfy the coding relation and zero otherwise. In Fig.~\ref{fig:factor_graph}, we use the $(5,7)_{\mathrm{oct}}$ convolutional code with rate $R=1/2$ and memory length $\nu=2$, giving $f_{\mathrm{en},1}(v_{k-2},v_k,c_{k,1})$ and $f_{\mathrm{en},2}(v_{k-2},v_{k-1},v_k,c_{k,2})$. For illustration, Nodes L to M and Nodes N to O show the individual coding constraints for the input windows $(v_7,v_8,v_9)$ at $k=9$ and $(v_{11},v_{12},v_{13})$ at $k=13$, respectively. The modulation and observation factors above the coded-bit variables are omitted for clarity.

\subsection{BP-Based Derivation for CL-SEC}

CL-SEC involves two product-form probability computations: constructing word-level probabilities from bit-level probabilities in \eqref{eq:wwscore} and fusing the physical- and application-layer distributions in \eqref{eq:cross_prob}. For a loop-free factor graph, the sum-product algorithm computes exact marginal distributions because the incoming BP messages at each variable or factor node are based on nonoverlapping subtrees and therefore do not reuse the same evidence \cite{Kschischang2001Factor}. Although the CL-SEC factor graph generally contains loops, we derive its product-form computations under the standard locally tree-like approximation, whereby the incoming BP messages are treated as if they originated from disjoint subtrees \cite{Kschischang2001Factor}. Source-bit interleaving supports this approximation by increasing loop lengths or separating the words involved, thereby weakening dependencies among incoming BP messages, as discussed in Section~\ref{sec:loop_alleviation}. Under this approximation, we now show that the CL-SEC probability computations follow from the BP message-update rules.

Consider the variable node representing a masked word at position $n\in\mathcal N_{\mathrm m}$, such as Node D in Fig.~\ref{fig:factor_graph}. The decoder-derived probabilities $P\big(u_{\tilde K_n+k}\big|\mathbf y\big),\ k=1,\ldots,K_n$, associated with the word's constituent bits, serve as the incoming physical-layer BP messages to the word-bit mapping factor at Node I. These bit probabilities are computed before word-level correction, i.e., without feedback from this stage, and therefore do not reuse information returned by the word-bit mapping factor. Let $\mu_{\mathrm{ph},n}(\tilde{\mathbf w}_{n,s})$ denote the downward factor-to-variable BP message (shown in magenta) from Node I to Node D for candidate $\tilde{\mathbf w}_{n,s}$. According to the sum-product update rule, this BP message is obtained by summing over all assignments of the associated bit variables:
\begin{align}
\mu_{\mathrm{ph},n}(\tilde{\mathbf{w}}_{n,s})
&= \sum_{\mathbf{b}_n\in\mathbb{F}_2^{K_n}}
f_{\mathrm{wb}}\big(\tilde{\mathbf{w}}_{n,s},\mathbf{b}_n\big) 
\prod_{k=1}^{K_n}
P\big(u_{\tilde{K}_n+k}=b_k \,\big|\, \mathbf{y}\big) \nonumber \\ 
&\stackrel{(a)}{=}\prod_{k=1}^{K_n}
P\big(u_{\tilde{K}_n+k}=\tilde{u}_{n,s}(k) \,\big|\, \mathbf{y}\big) \nonumber \\
&\stackrel{(b)}{\propto} d_{\mathrm{ph},n}(\tilde{\mathbf{w}}_{n,s}),\qquad s\in\mathcal{S}_n, 
\label{eq:bp_word_belief}
\end{align}
where $\mathbf b_n\triangleq(b_1,\ldots,b_{K_n})$. This BP message computation recovers the bit-to-word product-form computation in \eqref{eq:wwscore}, as shown by $(a)$ above, while $(b)$ follows from \eqref{eq:ph_distribution}.

Meanwhile, the semantic factor at Node A sends the application-layer BP message $\mu_{\mathrm{ap},n}(\tilde{\mathbf w}_{n,s})=\theta_n(\tilde{\mathbf w}_{n,s})$, shown in orange in Fig.~\ref{fig:factor_graph}, to the masked-word variable at Node~D. From \eqref{eq:theta} and \eqref{eq:ap_prob}, we have $\mu_{\mathrm{ap},n}(\tilde{\mathbf w}_{n,s})\propto d_{\mathrm{ap},n}(\tilde{\mathbf w}_{n,s})$. At Node~D, the sum-product rule multiplies the physical- and application-layer BP messages to obtain the final probability distribution of the word:
\begin{align}
\!\!d_{\mathrm{cl},n}(\tilde{\mathbf{w}}_{n,s})
&\propto\mu_{\mathrm{ph},n}(\tilde{\mathbf{w}}_{n,s}) \cdot \mu_{\mathrm{ap},n}(\tilde{\mathbf{w}}_{n,s}) \nonumber \\
&\propto d_{\mathrm{ph},n}(\tilde{\mathbf{w}}_{n,s}) \cdot d_{\mathrm{ap},n}(\tilde{\mathbf{w}}_{n,s}),\,
n\in\mathcal{N}_{\mathrm{m}},s\in\mathcal{S}_n. \label{eq:bp_cross_belief}
\end{align}
This gives the product-form cross-layer combination in \eqref{eq:cross_prob}.

\subsection{Interleaving-Enabled Loop Alleviation}\label{sec:loop_alleviation}

Loops can introduce dependencies among incoming BP messages, and thus applying the sum-product rules generally yields approximate distributions. Figure~\ref{fig:factor_graph}(a) contains many short loops that can introduce strong dependence among the combined BP messages. Source-bit interleaving, as illustrated in Fig.~\ref{fig:factor_graph}(b), can alleviate the effects of such loops by increasing their lengths or separating the words they connect. Although some short loops remain, interleaving can weaken the dependence among the combined BP messages overall, making their product-form combination in CL-SEC a more accurate approximation.\footnote{The classical BP algorithm also employs a combination of beliefs in product form, even when the corresponding factor graph contains loops. Nevertheless, this product-form approximation proves to be highly effective in many problems, such as LDPC decoding \cite{Kschischang2001Factor}.} We illustrate the effects of interleaving below. We denote a node labeled X in Fig.~\ref{fig:factor_graph} by $\mathrm{N}_{\mathrm{X}}$, where X may be a letter or a number. For numerical labels, $\mathrm N_k$ denotes the bit-variable node labeled $k$, which represents the encoder-input bit $v_k$.

\textbf{(i) Increasing the Loop Length:} In Fig.~\ref{fig:factor_graph}(a), the path
\begin{equation}\label{eq:loop_length_without_interleaving}
\mathrm{N}_{\mathrm{H}} - \mathrm N_{7} - \mathrm{N}_{\mathrm{M}} - \mathrm N_{8} - \mathrm{N}_{\mathrm{H}}
\end{equation}
forms a four-hop loop. Both bit-variable nodes $\mathrm N_{7}$ and $\mathrm N_{8}$ belong to the same word and connect to the same channel-coding factor. In Fig.~\ref{fig:factor_graph}(b), the bits occupying encoder-input positions 7 and 8 originate from different words. The corresponding loop through the semantic factor becomes
\begin{equation}\label{eq:loop_length_with_interleaving}
\mathrm{N}_{\mathrm{A}} - \mathrm{N}_{\mathrm{B}} - \mathrm{N}_{\mathrm{G}} - \mathrm N_{7} - \mathrm{N}_{\mathrm{M}}
- \mathrm{N}_8 - \mathrm{N}_{\mathrm{K}} - \mathrm{N}_{\mathrm{F}} - \mathrm{N}_{\mathrm{A}},
\end{equation}
which has eight hops. Thus, source-bit interleaving can replace a short loop within a word with a longer loop spanning different words, increasing the path length over which information can return to its point of origin.

\textbf{(ii) Separating the Involved Words:} Interleaving can also reduce the semantic dependence carried through a loop without changing its length. For example, Fig.~\ref{fig:factor_graph}(a) contains the eight-hop loop
\begin{equation}\label{eq:loop_separation_without_interleaving}
\mathrm{N}_{\mathrm{A}} - \mathrm{N}_{\mathrm{D}} - \mathrm{N}_{\mathrm{I}} - \mathrm N_{12} - \mathrm{N}_{\mathrm{O}}
- \mathrm{N}_{13} - \mathrm{N}_{\mathrm{J}} - \mathrm{N}_{\mathrm{E}} - \mathrm{N}_{\mathrm{A}},
\end{equation}
which connects the adjacent words represented by $\mathrm N_{\mathrm{D}}$ and $\mathrm N_{\mathrm{E}}$. In Fig.~\ref{fig:factor_graph}(b), the corresponding loop through the same nodes $\mathrm N_{12}$, $\mathrm N_{\mathrm{O}}$, and $\mathrm N_{13}$ is
\begin{equation}\label{eq:loop_separation_with_interleaving}
\mathrm{N}_{\mathrm{A}} - \mathrm{N}_{\mathrm{C}} - \mathrm{N}_{\mathrm{H}} - \mathrm N_{12} - \mathrm{N}_{\mathrm{O}}
- \mathrm{N}_{13} - \mathrm{N}_{\mathrm{K}} - \mathrm{N}_{\mathrm{F}} - \mathrm{N}_{\mathrm{A}}.
\end{equation}
Although this loop still has eight hops, it connects the more widely separated words at $\mathrm N_{\mathrm{C}}$ and $\mathrm N_{\mathrm{F}}$. For illustration purposes, Fig.~\ref{fig:factor_graph} shows only five words; for longer messages, a loop in the graph with interleaving can involve words that are much farther apart within the word sequence. In general, words within a local context tend to exhibit stronger linguistic dependencies than more distant words, and the MLM captures these contextual relationships through its learned representations. By connecting more widely separated words, source-bit interleaving can therefore weaken the semantic dependence carried through a loop, even without changing its length.

\textbf{(iii) Retaining Some Loops:} Consider in Fig.~\ref{fig:factor_graph}(a) the eight-hop loop
\begin{equation}\label{eq:loop_persistent_without_interleaving}
\mathrm{N}_{\mathrm{A}} - \mathrm{N}_{\mathrm{C}} - \mathrm{N}_{\mathrm{H}} - \mathrm N_{7} - \mathrm{N}_{\mathrm{L}}
- \mathrm{N}_9 - \mathrm{N}_{\mathrm{I}} - \mathrm{N}_{\mathrm{D}} - \mathrm{N}_{\mathrm{A}}.
\end{equation}
In Fig.~\ref{fig:factor_graph}(b), the corresponding loop through the same nodes $\mathrm N_{7}$, $\mathrm N_{\mathrm{L}}$, and $\mathrm N_{9}$ is
\begin{equation}\label{eq:loop_persistent_with_interleaving}
\mathrm{N}_{\mathrm{A}} - \mathrm{N}_{\mathrm{B}} - \mathrm{N}_{\mathrm{G}} - \mathrm N_{7} - \mathrm{N}_{\mathrm{L}}
- \mathrm{N}_9 - \mathrm{N}_{\mathrm{H}} - \mathrm{N}_{\mathrm{C}} - \mathrm{N}_{\mathrm{A}}.
\end{equation}
Both loops have eight hops and connect adjacent words. Therefore, in this example, source-bit interleaving leaves both the loop length and the separation between the connected words unchanged.

These examples illustrate that source-bit interleaving can mitigate short-loop effects by weakening dependencies among BP messages overall. The same reasoning applies to \textbf{block codes} such as LDPC codes. Within each code block, parity bits are generated according to the generator matrix as binary linear combinations of selected source bits, which may include nearby bits. The corresponding block-code constraints replace the convolutional-coding factors $f_{\mathrm{en}}$ in Fig.~\ref{fig:factor_graph}. Applying source-bit interleaving before block segmentation can disperse source bits that are close together in the original sequence across different code blocks. This can similarly mitigate short loops and local dependencies in the CL-SEC factor graph. The \textbf{ablation results} in Appendix~\ref{sec:supp} demonstrate improved CL-SEC performance with source-bit interleaving for both convolutional and LDPC codes.

\begin{remark}[Trellis-Based Exact BCJR Decoding]\label{rem:bcjr_trellis}
Figure~\ref{fig:factor_graph} uses a bit-level factorization of the convolutional-coding constraints to illustrate their connections with word-level constraints in CL-SEC. The same coding relations admit an equivalent state-based factorization, obtained by grouping the $\nu$ encoder-memory bits into a trellis state. Although the bit-level representation contains short loops, such as $\mathrm N_{7}-\mathrm N_{\mathrm{L}}-\mathrm N_{9}-\mathrm N_{\mathrm{M}}-\mathrm N_{7}$, the state-based factor graph is loop-free. BCJR performs BP on this latter representation and computes exact bit-wise posterior probabilities through forward-backward inference \cite{Kschischang2001Factor}.
\end{remark}

\begin{remark}[Interleaving and Graph Isomorphism]\label{rem:interleaving_isomorphism}
For both convolutional codes and block codes such as LDPC codes, source-bit interleaving changes the assignment of source bits to encoder-input positions while preserving the topology of the coding graph up to node relabeling. For example, the subgraphs formed by the top three rows of nodes in Figs.~\ref{fig:factor_graph}(a) and~\ref{fig:factor_graph}(b) are isomorphic: they have the same connectivity up to node relabeling and the same number of loops of each length. The isomorphism comparison here uses the same coding-graph representation before and after interleaving. Source-bit interleaving therefore preserves the loop structure of the chosen coding-graph representation: a loop-free graph remains loop-free, while in a loopy graph the number of loops $C(\zeta)$ of each length $\zeta$ remains unchanged. Our simulations further show no performance improvement from source-bit interleaving for BP-based standalone BCJR and LDPC decoding under the considered settings (Section~\ref{sec:sim0}).

When the bottom three rows of nodes in Fig.~\ref{fig:factor_graph} are further included, the word-level and semantic constraints introduce additional connections, and the graphs in Figs.~\ref{fig:factor_graph}(a) and~\ref{fig:factor_graph}(b) are no longer isomorphic. Source-bit interleaving can thus change the cross-layer loop structure or the words involved in these loops, as illustrated by the preceding examples.
\end{remark}

\begin{remark}[Masking for LM Inference]\label{rem:mlm_masking}
Figure~\ref{fig:factor_graph} also motivates masking for MLM-based contextual inference. Following the BP rule, the outgoing BP message from the semantic factor $f_{\mathrm{se}}$ to a word variable, such as the BP message $\mu_{\mathrm{ap},n}$ in orange, must exclude the incoming BP message on the same edge. Consistent with this principle, CL-SEC masks the current word position before MLM inference, so that its contextual prediction does not condition on the current word estimate.
\end{remark}

\section{Simulation Results}\label{sec:sim}

\subsection{Experimental Setup}\label{sec:sim0}

\textbf{Models:} We adopt mmBERT \cite{Marone2025mmBERT2}, a state-of-the-art MLM, as our model for masked-word inference in this section. We use Qwen3-4B \cite{Yang2025Qwen3} to restore punctuation with the prompt shown in Fig.~\ref{fig:punctuation}. To demonstrate CL-SEC's applicability across different models, Appendix~\ref{sec:supp} provides additional results using BART \cite{Lewis2020BART} as an alternative mask-infilling model.

\textbf{Dataset:} We use 150 passages as source messages for evaluation: 100 from general articles and 50 from daily emails, representing typical text payloads.\footnote{Data open available at: \textcolor{blue}{\url{https://github.com/Yirun719/CL-SEC}}.} Table~\ref{tab:dataset} summarizes the corpus statistics. We simulate the transmission of each passage under varying signal-to-noise ratio (SNR). This transmission procedure is repeated over 50 Monte-Carlo trials for each passage, yielding 7500 received passages at every tested SNR to obtain stable Monte-Carlo results.

\begin{table}[t]
\centering
\caption{Statistics of the corpus.}
\label{tab:dataset}
\begin{tabularx}{\columnwidth}{@{}lYYYY@{}}
\hline
& Minimum & Maximum & Mean & Std dev \\
\hline
Number of words & 71 & 207 & 112.9 & 32.8 \\
Number of letters & 342 & 1051 & 518.5 & 127.3 \\
\hline
\end{tabularx}
\end{table}

\textbf{System Settings:} We use both the convolutional code (CC) and LDPC code for evaluation. For CC, we apply the $(5,7)_{\mathrm{oct}}$ code with rate $R=1/2$ and memory length $\nu=2$, decoded by the BCJR algorithm. For LDPC, we employ a $(3,6)$-regular systematic code with $N_{\mathrm c}=720$ coded bits and $K_{\mathrm c}=362$ information bits per codeword, generated using the pyldpc library \cite{pyldpc}. Each payload is divided into blocks of $K_{\mathrm c}$ information bits, with zeros appended to the last block if needed. The resulting codewords are joined into a bitstream and rate matching is applied across the full payload to achieve an effective code rate of $R=1/2$. We use BP decoding for LDPC with a maximum of 80 iterations.

We consider both AWGN and Rayleigh fading channels, with the model in \eqref{eq:y}. Specifically, the noise is randomly sampled from $\mathbf n\sim\mathcal{CN}(\mathbf 0_Q,\sigma^2\mathbf I_Q)$. For Rayleigh fading, the channel coefficient is drawn independently from $\mathcal{CN}(0,1)$ for each transmission and remains constant over the payload. Perfect channel state information is assumed at the receiver for equalization. We use quadrature phase-shift keying (QPSK) with symbol amplitude $|x_q|=A,\,q=1,\ldots,Q$. The average received SNR for both channels is defined as
\begin{equation}
    \mathrm{SNR}=\frac{\mathbb{E}[|h x_q|^2]}{\sigma^2}=\frac{A^2}{\sigma^2}.
\end{equation}

\textbf{Investigated Schemes:} We compare the six schemes below. Schemes 1--5 are each evaluated with both CC and LDPC codes, whereas Scheme 6 is tailored to convolutional codes and is thus evaluated only with CC.

\textbf{Scheme 1: BCJR/BP Decoding.} This scheme applies BW-HD to the soft information produced by the BCJR or BP decoder (see \eqref{eq:u_hd}) and then converts the resulting bits into characters. The resulting word estimates may not be valid words. This scheme is essentially the classical channel decoding method.

\textbf{Scheme 2: WW-HD Scheme.} For each word position, this scheme constructs a physical-layer probability distribution over candidate words from the decoder's soft LLRs and selects the most probable candidate, as in \eqref{eq:wwscore}--\eqref{eq:w_ph}. It uses no LM inference.

\textbf{Scheme 3: MLM Scheme.} This scheme masks unreliable WW-HD estimates with confidence below a threshold, as in \eqref{eq:Nm}, and feeds the resulting partially masked sequence to the MLM. For each masked position, the MLM produces an application-layer probability distribution over a candidate set. The candidate word with the highest MLM probability, as in \eqref{eq:ap_prob}--\eqref{eq:app_corr}, is chosen to replace the mask. Unreliable WW-HD estimates are refined using only the application-layer MLM-derived distribution.\footnote{The standalone SEC approach in \cite{Hao2025Shortconf} also relies solely on the application-layer LM to refine post-FEC results. Specifically, it applies LM-based refinement to text reconstructed from conventional channel-decoding outputs, which may contain invalid words. Unlike Scheme 3, this approach does not employ the WW-HD mechanism of Scheme 2. Scheme 3 instead refines WW-HD estimates, thereby restricting the candidate set to valid words that satisfy the corresponding word-length constraints. Our simulations show that WW-HD produces more accurate preliminary word estimates than conventional channel decoding. We therefore use Scheme 3 as the representative benchmark for LM-only refinement initialized with these improved estimates. For clarity, we omit the curves corresponding to the approach in \cite{Hao2025Shortconf}.}

\textbf{Scheme 4: CL-SEC.} This scheme corresponds to the word-correction stage of the proposed framework. For each masked position with WW-HD confidence below a threshold, as in \eqref{eq:Nm}, this scheme combines the physical-layer and application-layer distributions and selects the candidate with the highest fused probability to replace the mask (see \eqref{eq:cross_prob}--\eqref{eq:cross_corr}).

\textbf{Scheme 5: CL-SEC\,+\,PR.} This scheme constitutes the complete proposed framework, applying Qwen3-based punctuation restoration (PR) to the CL-SEC results (Scheme 4).\footnote{Schemes 1--4 are compared without PR to assess their word-correction performance. Simulations show that CL-SEC (Scheme 4) outperforms Schemes 1--3. Additional experiments applying the same PR procedure to Schemes 1--4 show that CL-SEC\,+\,PR retains its performance advantage, consistent with the results before PR. For clarity, we omit the curves for Schemes 1--3 with additional PR.}

\textbf{Scheme 6: LLM-Viterbi.} LLM-Viterbi \cite{Li2026LLMViterbi2} incorporates semantic priors from a fine-tuned ByT5 model into Viterbi decoding for the convolutional code. It maintains multiple candidate paths and periodically ranks and prunes them using a metric that combines channel reliability with LM probabilities. It serves as an LM-assisted replacement for conventional Viterbi decoding, while the proposed framework refines the output of a soft channel decoder.

In our experiments, Schemes 2--5 use random source-bit interleaving before channel coding and deinterleaving after soft channel decoding. Scheme 1 serves as the conventional BCJR/BP baseline without this additional interleaving step (i.e., $\mathbf v=\mathbf u$). To validate, we have also evaluated source-bit interleaving for Scheme 1 and observed no performance improvement under the considered settings, as expected. LLM-Viterbi (Scheme 6) also operates without source-bit interleaving, consistent with its original formulation in \cite{Li2026LLMViterbi2}.

\textbf{Evaluation Metrics:}
We employ metrics for both transmission reliability and message fidelity. For transmission reliability, we report the bit error rate (BER), word error rate (WER), and character error rate (CER), which measure the verbatim recovery of the transmitted payload after excluding spaces and punctuation. For each SNR, we use the superscript $t\in\{1,\ldots,T\}$ to denote the individual transmission trial. At trial $t$, let $\hat u_k^{(t)}$, $\hat{\mathbf w}_n^{(t)}$, and $\hat s_\ell^{(t)}$ denote the estimates of source bit $u_k^{(t)}$, word $\mathbf w_n^{(t)}$, and character $s_\ell^{(t)}$, respectively, for $k=1,\ldots,K^{(t)}$, $n=1,\ldots,N^{(t)}$, and $\ell=1,\ldots,L^{(t)}$. Here, $K^{(t)}$, $N^{(t)}$, and $L^{(t)}$ are the total numbers of source bits, words, and characters, respectively, in the entire payload of trial $t$. Bit and character representations are related by the ASCII mapping. We aggregate the errors over all $T$ transmission trials as follows:
\begin{align}
    \mathrm{BER} &= \frac{1}{\sum_{t=1}^{T}K^{(t)}} \sum_{t=1}^{T}\sum_{k=1}^{K^{(t)}}\mathbb{I}\big(\hat{u}_k^{(t)}\ne u_k^{(t)}\big), \label{eq:eval_ber}\\
    \mathrm{WER} &= \frac{1}{\sum_{t=1}^{T}N^{(t)}} \sum_{t=1}^{T}\sum_{n=1}^{N^{(t)}}\mathbb{I}\big(\hat{\mathbf{w}}_n^{(t)}\ne\mathbf{w}_n^{(t)}\big), \label{eq:eval_wer}\\
    \mathrm{CER} &= \frac{1}{\sum_{t=1}^{T}L^{(t)}} \sum_{t=1}^{T}\sum_{\ell=1}^{L^{(t)}}\mathbb{I}\big(\hat{s}_\ell^{(t)}\ne s_\ell^{(t)}\big). \label{eq:eval_cer}
\end{align}

In addition, we assess the fidelity of each recovered message to its original message $\mathbf p$, including punctuation and spaces, using BERTScore \cite{Zhang*2020BERTScore}, ROUGE-L \cite{Lin2004ROUGE}, and BLEU \cite{Papineni2001BLEU}. BERTScore measures semantic similarity via contextual embeddings. ROUGE-L evaluates content preservation based on the longest common subsequence. BLEU evaluates lexical overlap using 1-gram through 4-gram precisions. Let $\hat{\mathbf p}^{(t)}$ and $\mathbf p^{(t)}$ be the recovered and original messages, respectively, at transmission trial $t$. For Schemes 1--4, $\hat{\mathbf p}^{(t)}$ is the recovered message with spaces between words but without punctuation. For Schemes 5--6, $\hat{\mathbf p}^{(t)}$ has both spaces and punctuation. At each SNR, the reported value of each metric is averaged over all $T$ transmission trials:
\begin{equation}\label{eq:eval_fidelity}
    F_{\mathrm{score}}=\frac{1}{T}\sum_{t=1}^{T}\operatorname{scorer}\big(\hat{\mathbf{p}}^{(t)},\mathbf{p}^{(t)}\big),
\end{equation}
where $\operatorname{scorer}\in\{\mathrm{BERTScore},\mathrm{ROUGE\text{-}L},\mathrm{BLEU}\}$ compares the recovered message with the original message. All message-fidelity scores are reported on a $[0,100]$ scale. As punctuation and spaces are excluded from the transmission-reliability calculations in \eqref{eq:eval_ber}--\eqref{eq:eval_cer}, Scheme 5 is only evaluated using message fidelities.

For Schemes 3--4, the masking threshold $\tau$ is calibrated offline. Specifically, we consider each combination of correction scheme, channel code, channel model, and SNR separately and select the threshold that minimizes BER, a representative metric of verbatim recovery. The resulting thresholds are stored in a lookup table for online selection based on the known system configuration and the channel condition estimated at the receiver. This approach is analogous to parameter selection in adaptive modulation and coding in modern communication systems. It requires no additional signaling or channel-estimation overhead, and the online operation involves only a table lookup with negligible computational cost.

\begin{figure*}[!t]
    \centering
    \includegraphics[width=0.90\linewidth] {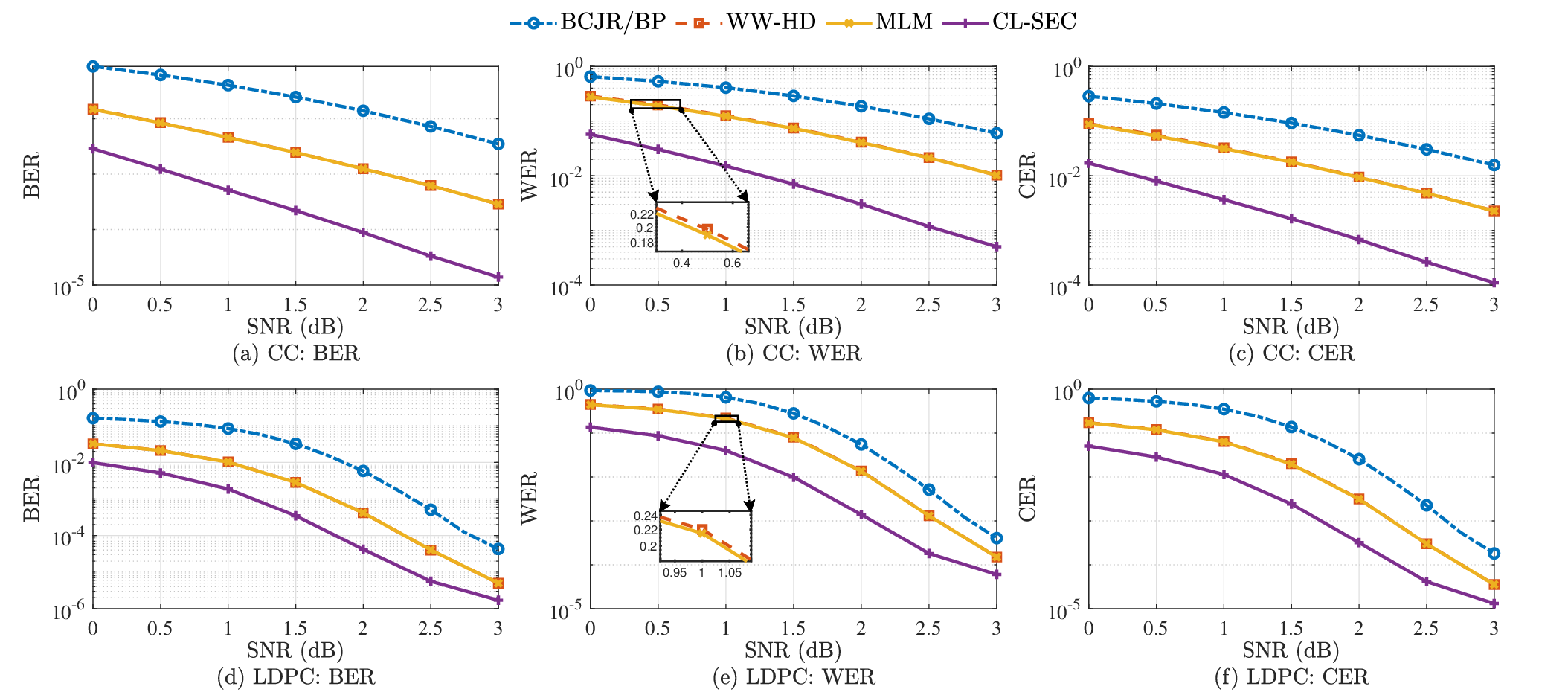} 
    \caption{Component-wise comparison over AWGN: (a) BER, (b) WER, and (c) CER for CC; (d) BER, (e) WER, and (f) CER for LDPC.}
    \label{fig:awgn-bwcer}
    \par\vspace{\floatsep}
    \includegraphics[width=0.90\linewidth]{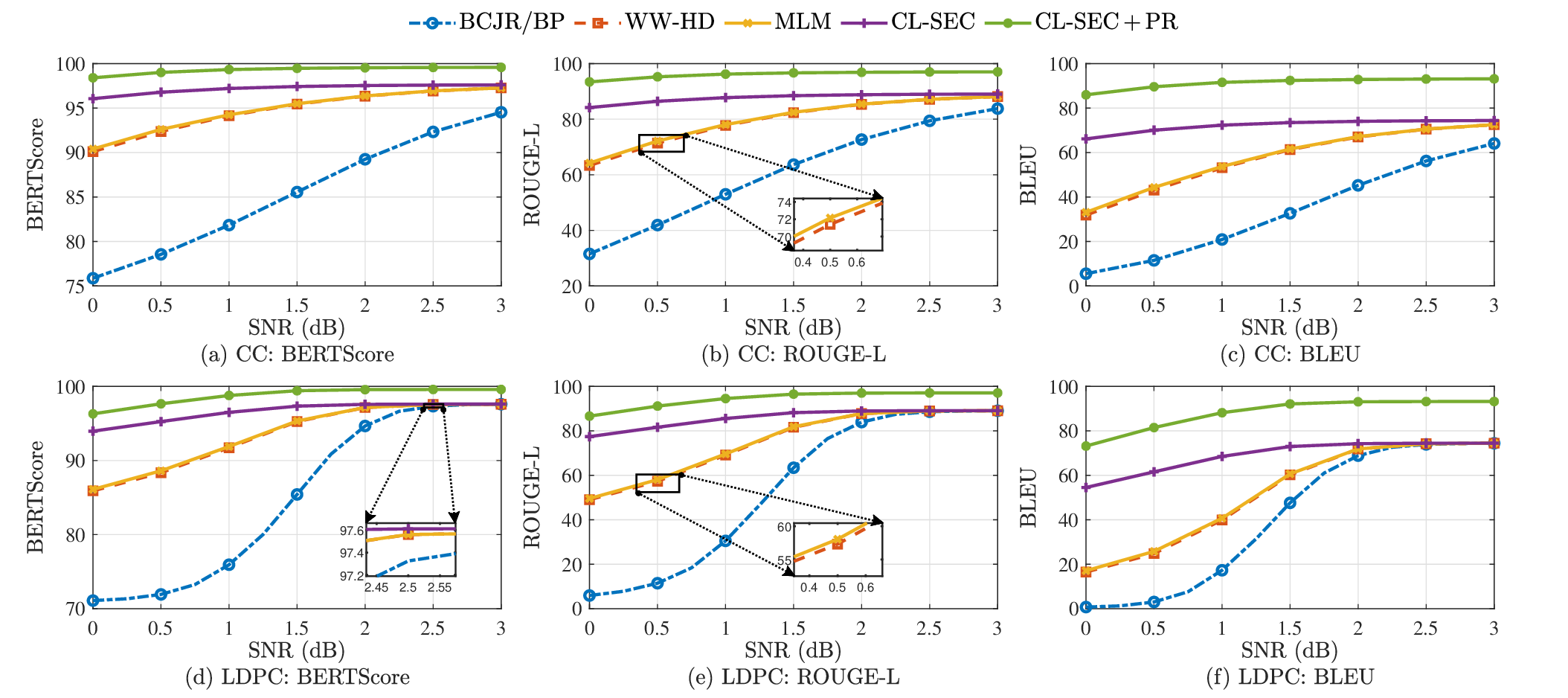}
    \caption{Component-wise comparison over AWGN: (a) BERTScore, (b) ROUGE-L, and (c) BLEU for CC; (d) BERTScore, (e) ROUGE-L, and (f) BLEU for LDPC.}
    \label{fig:awgn-semantic}
\end{figure*}

\subsection{Component-Wise Comparison over AWGN Channels}

This subsection compares Schemes 1--5 using both CC and LDPC over AWGN channels. Figures~\ref{fig:awgn-bwcer} and~\ref{fig:awgn-semantic} report the transmission-reliability and message-fidelity results, respectively.

The BER results are presented in Figs.~\ref{fig:awgn-bwcer}(a) and~\ref{fig:awgn-bwcer}(d). The classical BER results of channel decoding---BCJR for CC and BP for LDPC---serve as important physical-layer benchmarks for the other schemes, which incorporate different degrees of semantic information in addition to physical-layer signal information.

By incorporating semantic information in a principled manner, CL-SEC significantly improves the conventional BER performance of FEC decoding and achieves superior verbatim recovery. For example, at an SNR of 2\,dB, CL-SEC reduces the BER from $1.39\times10^{-2}$ to $8.83\times10^{-5}$ for CC and from $5.84\times10^{-3}$ to $4.19\times10^{-5}$ for LDPC, corresponding to approximately $157$-fold and $139$-fold reductions relative to conventional BCJR and BP decoding, respectively.

Moreover, CL-SEC consistently outperforms WW-HD for both codes, demonstrating the effectiveness of refining unreliable WW-HD estimates through cross-layer fusion.
At an SNR of 1\,dB, CL-SEC reduces the BER relative to WW-HD by factors of approximately $9.01$ and $5.49$ for CC and LDPC, respectively. In contrast, MLM-only refinement provides only marginal gains. At an SNR of 1\,dB, it reduces the BER from $4.626\times10^{-3}$ to $4.571\times10^{-3}$ for CC and from $1.027\times10^{-2}$ to $1.017\times10^{-2}$ for LDPC, corresponding to reductions of only $1.19\%$ and $0.97\%$, respectively. These results show that using MLM predictions alone to refine unreliable WW-HD estimates provides limited benefits, whereas fusing soft LLR evidence with MLM probabilities in CL-SEC substantially improves correction accuracy.

In addition, WW-HD outperforms conventional BCJR/BP decoding followed by BW-HD. At an SNR of 1.5\,dB, WW-HD reduces the BER by factors of approximately $9.96$ and $11.28$ for CC and LDPC, respectively. These results demonstrate the effectiveness of WW-HD in converting the decoder's soft information into hard decisions. The WER and CER results in Fig.~\ref{fig:awgn-bwcer} exhibit trends consistent with those observed for BER.

The three fidelity metrics in Fig.~\ref{fig:awgn-semantic} show similar trends, further supporting the effectiveness of the proposed framework in message recovery. CL-SEC substantially improves message fidelity over conventional channel decoding at low SNRs, and PR provides additional gains across all three fidelity metrics. At an SNR of 0\,dB, CL-SEC\,+\,PR achieves BERTScores of $98.41$ and $96.29$ for CC and LDPC, respectively, demonstrating reliable recovery of semantic meaning even under low-SNR conditions.

\begin{figure*}[!t]
    \centering
    \includegraphics[width=0.875\linewidth]{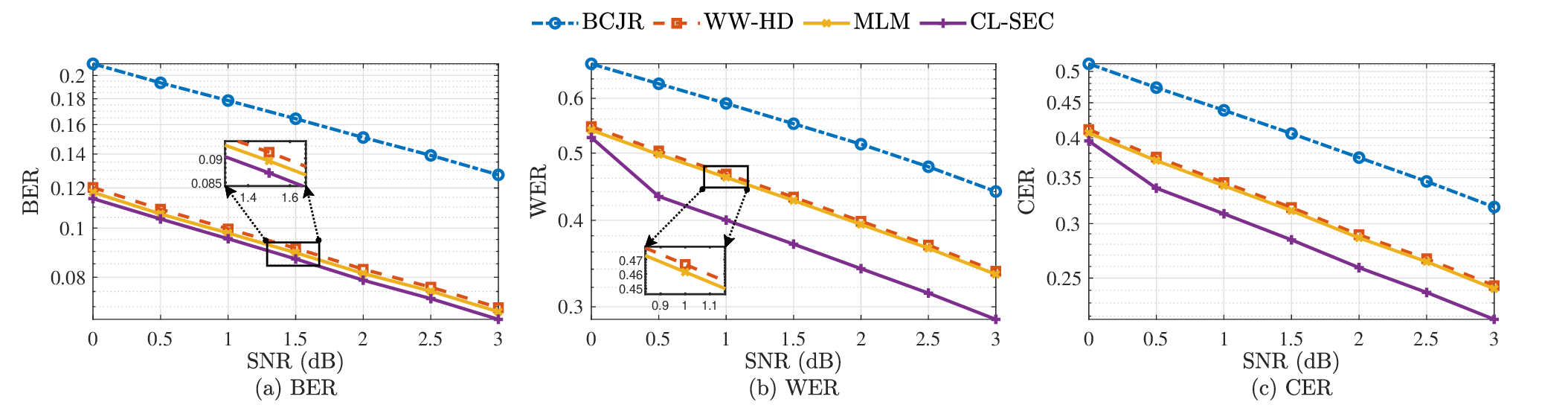}
    \caption{Component-wise comparison using CC over Rayleigh fading: (a) BER, (b) WER, and (c) CER.}
    \Description{Rayleigh-fading verbatim-recovery results}
    \label{fig:rayleigh-bwcer}
    \par\vspace{\floatsep}
    \includegraphics[width=0.875\linewidth]{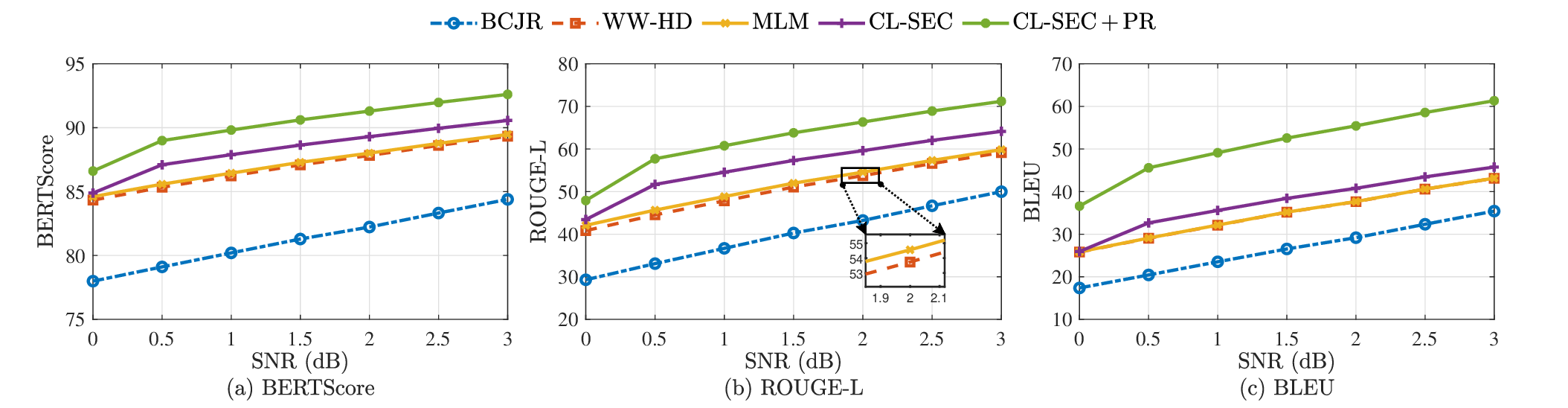}
    \caption{Component-wise comparison using CC over Rayleigh fading: (a) BERTScore, (b) ROUGE-L, and (c) BLEU.}
    \Description{Rayleigh-fading semantic-fidelity results}
    \label{fig:rayleigh-semantic}
\end{figure*}

\subsection{Component-Wise Comparison over Rayleigh Fading}

This subsection compares Schemes 1--5 using CC over Rayleigh fading channels. Figure~\ref{fig:rayleigh-bwcer} reports the transmission-reliability results, while Fig.~\ref{fig:rayleigh-semantic} presents the message-fidelity results. The corresponding LDPC results exhibit similar trends and are omitted for brevity.

Compared with the AWGN results in Figs.~\ref{fig:awgn-bwcer} and~\ref{fig:awgn-semantic}, Rayleigh fading generally degrades recovery performance, as deep-fading events can substantially reduce the instantaneous received SNR, severely deteriorating the average recovery performance. Nevertheless, CL-SEC remains superior under fading, consistently outperforming BCJR, WW-HD, and MLM schemes across all considered metrics. These results demonstrate the robustness of cross-layer fusion under challenging channel conditions.

The BER, WER, and CER results in Fig.~\ref{fig:rayleigh-bwcer} exhibit consistent trends. Taking WER at $\operatorname{SNR}=1$\,dB as an example, CL-SEC reduces the WER from $0.590$ for BCJR to $0.400$, a relative reduction of $32.20$\%. Relative to WW-HD, whose WER is $0.467$, CL-SEC achieves a $14.35$\% reduction. In contrast, the MLM-only refinement reduces the WW-HD WER by only $1.28$\%, to $0.461$. These results show that fusing contextual predictions with soft LLRs substantially improves performance over both FEC decoding and WW-HD, whereas the MLM-only refinement provides limited gains. These findings are consistent with the AWGN results.

The message-fidelity results in Fig.~\ref{fig:rayleigh-semantic} further demonstrate the effectiveness of the proposed framework in recovering the original message. At an SNR of 2\,dB, CL-SEC improves the ROUGE-L score by $37.90\%$ over BCJR and by $10.92\%$ over WW-HD. Applying PR further increases the ROUGE-L score of CL-SEC by $11.29$\%. Moreover, CL-SEC\,+\,PR achieves BERTScores above $90$ at $\operatorname{SNR}\ge1.5$\,dB over Rayleigh fading channels. Across the tested SNRs, CL-SEC\,+\,PR also achieves an average relative BLEU improvement of 97.85\% over BCJR.

\begin{figure*}[!t]
    \centering
    \includegraphics[width=0.875\linewidth]{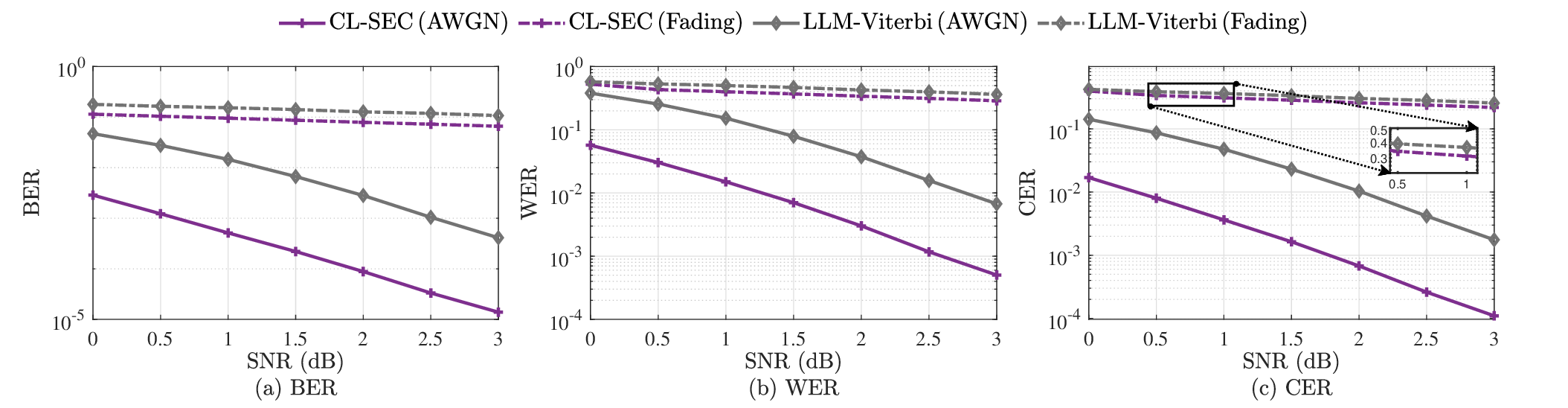}
    \caption{Comparison between CL-SEC and LLM-Viterbi using CC over AWGN and Rayleigh fading: (a) BER, (b) WER, and (c) CER.}
    \Description{Verbatim-recovery comparison with LLM-Viterbi}
    \label{fig:benchmark-bwcer}
    \par\vspace{\floatsep}
    \includegraphics[width=0.875\linewidth]{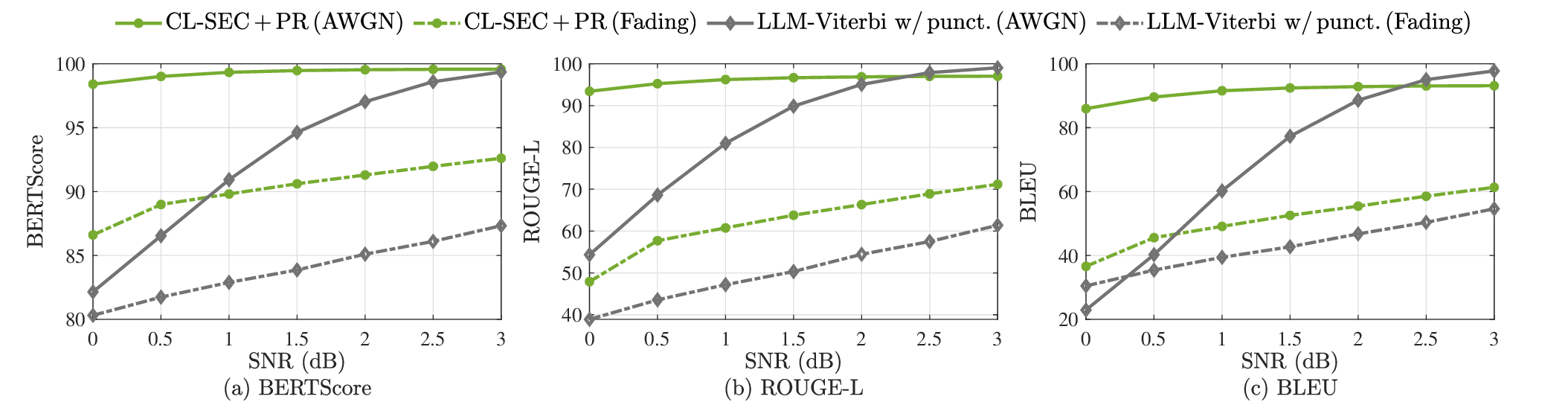}
    \caption{Comparison between CL-SEC\,+\,PR and LLM-Viterbi with punctuation (w/ punct.) using CC over AWGN and Rayleigh fading: (a) BERTScore, (b) ROUGE-L, and (c) BLEU.}
    \Description{Semantic-fidelity comparison with LLM-Viterbi}
    \label{fig:benchmark-semantic}
\end{figure*}

\subsection{Comparison with LLM-Viterbi}\label{sec:llmv-comparison}

We compare the proposed framework (CL-SEC or CL-SEC\,+\,PR) with LLM-Viterbi over both AWGN and Rayleigh fading channels. CL-SEC (Scheme 4) corresponds to the word-correction stage of the proposed framework, while CL-SEC\,+\,PR (Scheme 5) completes the proposed framework by further restoring punctuation. 
Since LLM-Viterbi is tailored to convolutional codes, we use the CC-based CL-SEC for a fair comparison. 
Figure~\ref{fig:benchmark-bwcer} compares CL-SEC (Scheme 4) with LLM-Viterbi in terms of transmission-reliability metrics. These metrics are computed based on \eqref{eq:eval_ber}--\eqref{eq:eval_cer} without punctuation.

Figure~\ref{fig:benchmark-semantic} compares CL-SEC\,+\,PR (Scheme 5) and LLM-Viterbi by evaluating the message-fidelity metrics. These metrics are computed using \eqref{eq:eval_fidelity}, with punctuation included in both the recovered and original messages. Unlike CL-SEC\,+\,PR, which strips punctuation during transmission and restores it at the receiver post-decoding, LLM-Viterbi transmits punctuation directly within the message to be decoded jointly at the receiver. The label ``LLM-Viterbi w/ punct.'' in Fig.~\ref{fig:benchmark-semantic} indicates that the decoded punctuation is included in the recovered message when scoring fidelity for LLM-Viterbi.

Figure~\ref{fig:benchmark-bwcer} shows that CL-SEC consistently outperforms LLM-Viterbi across the tested SNRs under both channel models. The gains are particularly substantial over AWGN channels. At an SNR of 1\,dB, CL-SEC reduces BER, WER, and CER by $96.49\%$, $90.15\%$, and $92.38\%$, respectively, relative to LLM-Viterbi. Under Rayleigh fading, the corresponding reductions are $37.91\%$, $20.21\%$, and $14.52\%$. Although the relative gains under Rayleigh fading are smaller, CL-SEC maintains an advantage across all three reliability metrics.

Figure~\ref{fig:benchmark-semantic} further shows that, over AWGN channels, CL-SEC\,+\,PR substantially outperforms ``LLM-Viterbi w/ punct.'' at low SNRs, although the latter achieves slightly higher fidelity scores at SNRs of 2.5\,dB and above. Under Rayleigh fading, both schemes generally achieve lower message-fidelity scores than under AWGN channels. Nevertheless, CL-SEC\,+\,PR consistently outperforms its counterpart across all three fidelity metrics, demonstrating the robustness of the proposed framework under fading conditions.

\section{Related Work}\label{sec:related}

The rapid advancement of language models (LMs) has opened new avenues for error-correction techniques. A recent study \cite{Hao2025Shortconf} proposed an LM-based semantic error correction (SEC) approach that complements conventional forward error correction (FEC). After channel decoding, it converts the decoded bits into text and passes this character sequence to the encoder of a fine-tuned BART model. The BART decoder then generates a revised character sequence autoregressively, conditioning each prediction on the encoded input and the output generated so far. This allows it to revise any part of the input text. This scheme is a standalone SEC approach that uses the LM for error correction at the application layer. The study \cite{Hao2025Shortconf} reported that LM-based SEC improves both semantic fidelity and block error rate (BLER).\footnote{Note that BLER is a metric for verbatim recovery---a character block is deemed wrong even with a single incorrect character. Prior standalone SEC schemes \cite{Hao2025Shortconf}, \cite{Guo2025Semantic}, \cite{Qiao2025Tokendomainconf} do not incorporate physical-layer soft reliability information during semantic error correction and tend to be inferior on verbatim-recovery metrics such as BLER here.}

Another study \cite{Guo2025Semantic} also employed standalone SEC after channel decoding. The receiver uses an LM to infer corrupted words based on the surrounding text, thereby reducing semantic loss. In addition, \cite{Qiao2025Tokendomainconf} investigated a non-orthogonal multiple-access framework with standalone SEC at the receiver. Simultaneous transmission of identical coded symbols by multiple users can prevent the receiver from decoding the data at that specific position. The system then masks these undecodable positions and uses an LM to predict the missing symbols based on the surrounding context.

Overall, the existing standalone SEC schemes \cite{Hao2025Shortconf}, \cite{Guo2025Semantic}, \cite{Qiao2025Tokendomainconf} operate exclusively at the application layer after hard decisions have been made at the physical layer, and therefore they rely solely on the application-layer information. Consequently, they overlook the valuable physical-layer soft information that can support error-correction refinement, particularly bit-level soft log-likelihood ratios. By relying on a single source of information---semantic context---these standalone schemes remain suboptimal and may suffer from \emph{semantic inaccuracy} and \emph{lexical imprecision}, as discussed in Section~\ref{sec:intro}.

To address these limitations, our work leverages soft information from both the physical and application layers to refine the output of FEC decoding. The proposed Cross-Layer SEC (CL-SEC) framework exploits the semantic and linguistic capabilities of LMs while retaining physical-layer soft information, with the objective of recovering the transmitted message \textbf{verbatim}.\footnote{We note that LLM-Viterbi \cite{Li2026LLMViterbi2} also combines physical- and application-layer information to improve conventional Viterbi decoding for convolutional codes. Its first arXiv version appeared after our initial CL-SEC preprint \cite{Wang2026CLSEC}. Specifically, LLM-Viterbi periodically scores and prunes the candidate trellis paths based on a joint score computed from both channel reliability and LM probabilities. The simulation results in Section~\ref{sec:llmv-comparison} show substantial gains achieved by CL-SEC relative to LLM-Viterbi.}

In many semantic communication systems, the primary objective is to preserve the meaning of the original message rather than achieve verbatim recovery, which is typically required in conventional communication systems. Verbatim recovery is crucial in applications requiring exact message integrity, such as the transmission of protocol files, legal documents, financial contracts, and scientific data, where even minor alterations can cause significant misunderstandings or errors. The system presented in this paper focuses on verbatim message recovery and leverages the semantic understanding of LMs to achieve this objective.

Building on the CL-SEC framework introduced in this work (an earlier version of which was released as a preprint \cite{Wang2026CLSEC}), a subset of the authors presented a two-page demonstration \cite{Pan2026Demo}. This work implemented CL-SEC on a software-defined radio (SDR) testbed using SDR middleware to provide an experimental proof of concept for real-time feasibility over actual wireless links. While \cite{Pan2026Demo} focused strictly on system implementation, the present paper comprehensively develops the CL-SEC framework itself, serving as the underlying theoretical foundation of \cite{Pan2026Demo}.

Finally, it is important to clarify that SEC---the overarching framework encompassing the aforementioned literature and our own work---is distinct from joint source-channel coding (JSCC) \cite{Wu2024DeepJSCCMIMO,Shao2023PAPR,Wu2024Transformer,Bian2023DeepJSCC,Xie2021DeepSC}. While both paradigms leverage artificial intelligence (AI) techniques to optimize communication systems, they do so through fundamentally different methodologies. Whereas JSCC replaces part or all of the conventional source- and channel-coding pipeline with an end-to-end learned transceiver, SEC can retain the standard channel-coding architecture and augment the receiver with an AI-empowered correction module \cite{Yue2026SECC}. It can therefore preserve compatibility with existing physical-layer designs while providing a practical pathway for AI to improve end-to-end communication reliability.

\section{Conclusion and Future Work}\label{sec:conclusion}

Shannon's channel coding theorem establishes the fundamental limit of reliable communication over a specified channel. Over the decades, advances in coding and decoding have enabled modern communication systems to approach this limit. However, conventional channel decoders generally do not exploit the linguistic structure and semantic context inherent in text. Recent advances in language models (LMs) provide a powerful means of leveraging such source-level information to complement physical-layer decoding and improve end-to-end message recovery.

In this work, we introduced CL-SEC, a Cross-Layer Semantic Error-Correction framework that integrates physical-layer soft information with application-layer semantic context to improve the \textbf{verbatim recovery} of transmitted text. Unlike conventional channel decoding, which does not explicitly exploit linguistic or semantic context, and existing SEC approaches, which generally do not incorporate bit-level soft information, CL-SEC jointly leverages information from both layers.

CL-SEC uses words as its basic correction units and derives physical-layer candidate distributions from the bit-wise soft information produced by the channel decoder. Reliable word estimates are retained, whereas unreliable estimates are masked and refined using a combination of physical-layer candidate probabilities and LM-derived application-layer probabilities. The framework further employs context-based punctuation restoration to reconstruct complete messages. We have established a belief-propagation-based factor-graph foundation for the product-form probability computations and introduced source-bit interleaving to mitigate short-loop effects. CL-SEC is compatible with different soft-output channel decoders and language models, allowing it to accommodate diverse communication conditions.

Experiments using convolutional and LDPC codes over AWGN and Rayleigh fading channels demonstrated that CL-SEC consistently improves bit error rate, word error rate, character error rate, and message fidelity scores compared with conventional channel decoding and methods that process physical- and application-layer information separately. These results highlight the benefits of jointly exploiting decoder soft information and linguistic context for accurate message reconstruction.

Our work has demonstrated that source-level linguistic and semantic information can provide substantial improvements beyond those achieved by channel decoding alone. Bridging physical-layer decoding and semantic error correction may therefore open new directions in communication-system design, in which source structure and meaning complement physical-layer information to enable more reliable transmission. 
Several avenues for future research remain:
\begin{itemize}[leftmargin=*, label=$\bullet$]
    \item \textbf{Source Compression:} 
    The current framework does not apply source compression to the transmitted payload. Future work can investigate the tradeoff between compression efficiency and the residual source redundancy available for semantic error correction under a fixed transmission budget.
    \item \textbf{Theoretical Framework:} 
    The belief-propagation-based formulation developed in this work provides a starting point for further theoretical analysis. Future research can build on this formulation to characterize the performance and fundamental limits of semantic error correction under explicit source and channel models.
    \item \textbf{Word-Length Inference:} 
    The current implementation includes word-length metadata in the frame header to constrain the candidate correction space. Future work can develop techniques for accurately inferring word lengths without relying on explicit metadata.
    \item \textbf{Processing Latency:} Refinement following channel decoding, particularly LM-based inference, introduces additional computational overhead and latency relative to conventional channel decoding. Future work can investigate efficient inference and implementation techniques to reduce this overhead for latency-sensitive applications.
\end{itemize}	

\appendices

\section{Canonical Huffman Coding for Word-Length Information Compression}\label{sec:header}

\begin{table*}[t]
    \centering
    \caption{Word-length statistics and the resulting canonical Huffman codebook.}
    \label{tab:huffman}
    \begin{tabularx}{\textwidth}{l|*{12}{>{\centering\arraybackslash}X}}
        \hline
        Word length (in letters) & 3 & 4 & 2 & 5 & 6 & 8 & 7 & 10 & 1 & 9 & 12 & 11 \\
        \hline
        Word count & 31 & 21 & 19 & 14 & 10 & 8 & 7 & 4 & 4 & 2 & 1 & 0 \\
        Canonical Huffman codeword & 00 & 01 & 100 & 101 & 1100 & 1110 & 1101 & 11110 & 111110 & 1111110 & 1111111 & None \\
        Codeword length (in bits) & 2 & 2 & 3 & 3 & 4 & 4 & 4 & 5 & 6 & 7 & 7 & ``0'' \\
        \hline
    \end{tabularx}
\end{table*}

\begin{table}[t]
    \centering
    \caption{Statistics of the header overhead.}
    \label{tab:overhead}
    \begin{tabular}{lcccc}
        \hline
        & Minimum & Maximum & Mean & Std dev \\
        \hline
        Overhead $\rho$ & 7.69\% & 10.77\% & {\bf8.71}\% & 0.45\% \\
        \hline
    \end{tabular}
\end{table}

To reduce overhead of the word-length metadata, we employ Huffman coding \cite{sayood2017introduction}. Huffman coding is a lossless compression algorithm that assigns shorter binary codewords to more frequent symbols and longer codewords to less frequent ones, thereby minimizing the average codeword length. In the basic Huffman coding framework, the entire Huffman tree must be transferred to the decoder, resulting in significant codebook overhead. Canonical Huffman coding is a standardized variant that reorganizes codewords such that only their lengths need to be transmitted. This canonical form enables deterministic reconstruction of the codebook at the decoder while reducing codebook overhead. 

In this work, we employ canonical Huffman coding to efficiently compress the word-length information of the message. The compressed word lengths are embedded in the frame header as metadata. For each word length $L_n,\,n\in\mathcal{N}$ in the message, we denote the length of its associated canonical Huffman codeword by $\Theta_n$ (in bits). Thus, we require $\sum_{n\in\mathcal{N}} \Theta_n$~bits to transmit word lengths. In addition, for codebook delivery, transmitting the length of each canonical Huffman codeword suffices, as mentioned earlier. We use fixed-length binary coding for codebook delivery and assume such delivery requires additional $\Phi$ bits. The codeword lengths associated with each word length can be transmitted in a preset word-length order. The decoder can thus rely on this order to determine each pair of codeword length and word length, based on which the decoder can further construct the complete canonical Huffman codebook uniquely. Relative to the payload $\sum_{n\in\mathcal{N}} K_n$~bits for word transmission, the header overhead $\rho$ is defined as
\begin{equation}\label{eq:huffman}
    \rho = \frac{\sum_{n\in\mathcal{N}} \Theta_n + \Phi}{\sum_{n\in\mathcal{N}} K_n}.
\end{equation}

To illustrate coding details and the compression efficiency, consider the following example, where the message is: 
\begin{tcolorbox}[
    colback=gray!8,        
    colframe=black,        
    boxrule=0.3pt,         
    arc=1pt,               
    breakable,             
    left=1pt,              
    right=1pt,             
    top=1pt,               
    bottom=1pt             
    ]
    \footnotesize\ttfamily  
    Modern climbers try to climb mountains by a route which will give them good sport, and the more difficult it is, the more highly it is regarded. In the pioneering days, however, this was not the case at all. The early climbers were looking for the easiest way to the top, because the summit was the prize they sought, especially if it had never been attained before. It is true that during their adventures they often faced difficulties and dangers of the most perilous nature, equipped in a manner which would make modern climbers shudder at the thought, but they did not go out of their way to court such excitement. They had a single aim, a solitary goal \textnormal{\textendash} the top!
\end{tcolorbox}
\noindent After removing punctuation and spaces, the message contains $N = 121$ words, corresponding to $\sum_{n\in\mathcal{N}} K_n = 4256$~bits in the payload. The word-length statistics and the resulting canonical Huffman codebook are listed in Table~\ref{tab:huffman}. With this codebook, transmitting $N$ word lengths in this message requires $\sum_{n\in\mathcal{N}} \Theta_n = 368$~bits. Regarding codebook delivery, we represent the length of each Huffman codeword using a fixed-length binary code. Note that there is no word in this message with word length 11. We reserve codeword length 0 for such a nonexistent word length. Consequently, we encode the codeword lengths varying from the minimum 0 to the maximum 7 using 3 bits, covering lengths $0, 1, \ldots, 7$. Thus, transmitting 11 codeword lengths and a reserved codeword length 0 requires an additional $\Phi = 36$~bits, which is much smaller than $\sum_{n\in\mathcal{N}} \Theta_n=368$~bits for word-length transmission. The codeword lengths corresponding to different word lengths can be transmitted in a preset order, e.g., in the ascending word-length order. With this order, the decoder can determine the codeword length for each word length, and can further construct the canonical Huffman codebook uniquely. Summing two parts of the header overhead, the total overhead for word-length transmission becomes $\sum_{n\in\mathcal{N}} \Theta_n + \Phi = 404$~bits. Relative to the payload $\sum_{n\in\mathcal{N}} K_n = 4256$~bits, the header overhead $\rho $ is 9.49\% using \eqref{eq:huffman}. This overhead is small in practical systems.

\begin{figure*}[!t]
    \centering
    \includegraphics[width=0.865\linewidth]{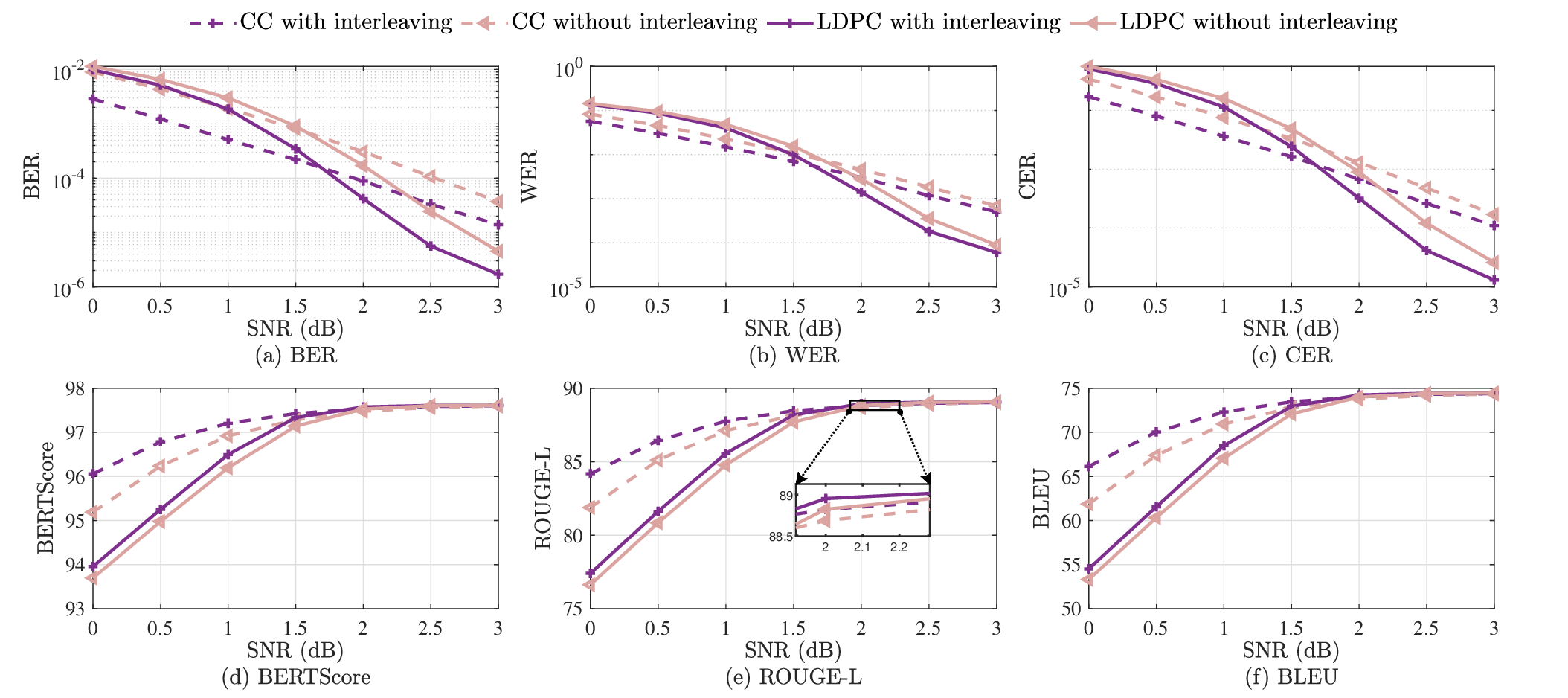}
    \caption{Effect of source-bit interleaving on CL-SEC performance using CC and LDPC codes over AWGN: (a) BER, (b) WER, (c) CER, (d) BERTScore, (e) ROUGE-L, and (f) BLEU.}
    \Description{Source-bit-interleaving ablation}
    \label{fig:interleaving}
    \par\vspace{\floatsep}
    \includegraphics[width=0.865\linewidth]{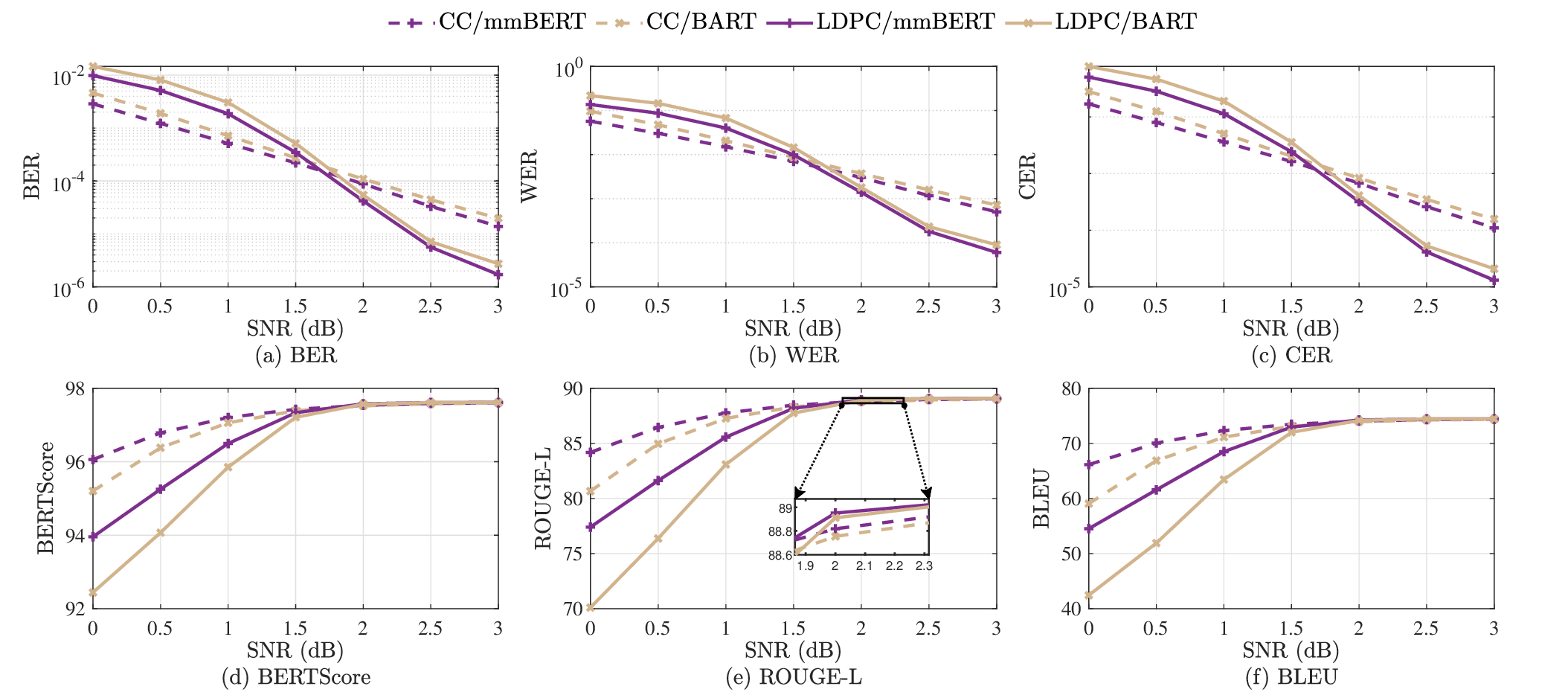}
    \caption{Performance comparison of mmBERT- and BART-based CL-SEC using CC and LDPC codes over AWGN: (a) BER, (b) WER, (c) CER, (d) BERTScore, (e) ROUGE-L, and (f) BLEU.} 
    \Description{Comparison of the mmBERT and BART implementations}
    \label{fig:mlm-generality}
\end{figure*}


The compression efficiency is further extensively evaluated over a message corpus (including the message example above).
For diverse evaluation, we include 650 passages in this corpus, consisting of the 150 passages specified in Section~\ref{sec:sim0} and the 500 passages from the Brown Corpus. 
The statistics of the header overhead are listed in Table~\ref{tab:overhead}. Across the corpus, {\bf98.77\%} of the passages incur an overhead of {\bf less than 10\%}, indicating that the header overhead is small in most practical cases.

\section{Supplementary Simulation Results}\label{sec:supp}
This appendix provides additional results on the effectiveness of source-bit interleaving and the applicability of CL-SEC across different LMs. 
The simulation settings follow those specified in Section~\ref{sec:sim0}. Results in this appendix are obtained using Scheme~4 (CL-SEC).

\textbf{Ablation Study on Source-Bit Interleaving:}
Section~\ref{sec:justification} explains how source-bit interleaving reduces the prevalence of short loops in the CL-SEC factor graph, thereby reducing their adverse effects on probability computation. To evaluate the effectiveness of the interleaving, Fig.~\ref{fig:interleaving} compares the performance of CL-SEC with and without source-bit interleaving over AWGN channels. 

Source-bit interleaving consistently improves all six performance metrics for both CC and LDPC codes. For example, when averaged over the considered SNRs, the relative BER reductions are {\bf69.83\%} for CC and 
{\bf50.05\%} for LDPC. These results demonstrate the effectiveness of source-bit interleaving and support the factor-graph justification in Section~\ref{sec:justification}: reducing the prevalence of short loops improves the accuracy of the product-form probability approximations employed by CL-SEC.

\textbf{Applicability of CL-SEC Across LMs:}
To investigate whether CL-SEC can operate with models other than mmBERT, we adopt BART \cite{Lewis2020BART} as an alternative mask-infilling model. Figure~\ref{fig:mlm-generality} presents the performance of BART-based CL-SEC over AWGN channels. BART-based CL-SEC achieves substantial performance improvements relative to conventional channel decoding (Figs.~\ref{fig:awgn-bwcer} and \ref{fig:awgn-semantic}) with both CC and LDPC codes, demonstrating that CL-SEC is not limited to mmBERT. However, Fig.~\ref{fig:mlm-generality} shows that mmBERT-based CL-SEC outperforms its BART-based counterpart across the evaluated metrics under the considered simulation settings. We therefore adopt mmBERT as the primary language model in this paper.


\bibliographystyle{IEEEtran}
\bibliography{ref.bib}

\end{document}